\documentclass[11pt,a4paper]{article}
\usepackage{amsmath,graphicx,amssymb,slashed,amsfonts,cancel,flushend}
\RequirePackage[T1]{fontenc}
\usepackage[utf8x]{inputenc}
\RequirePackage[numbers,sort&compress]{natbib}
\usepackage{authblk}
\usepackage[colorlinks,citecolor=blue,urlcolor=blue,linkcolor=blue]{hyperref}

\usepackage[a4paper, inner=2.5cm, outer=2.5cm, top=2.5cm,bottom=2.5 cm, binding offset=0 cm]{geometry}
\usepackage[toc]{appendix}
 
\def\MGvATNLO{ {\tt {\sc MadGraph5}\_aMC@NLO}}

\def\wtil#1{\widetilde{#1}}

\title{Anomalous triple gauge boson couplings in $ZZ$ production at the LHC  and  the role of $Z$ boson polarizations}
\author{Rafiqul Rahaman\thanks{rr13rs033@iiserkol.ac.in} }
\author{Ritesh K. Singh\thanks{ritesh.singh@iiserkol.ac.in} }
\affil{Department of Physical Sciences,
	Indian Institute of Science Education and Research Kolkata,
	Mohanpur, 741246, India}
\date{}
\begin{document}
	\maketitle

\begin{abstract}
We study anomalous couplings among neutral gauge bosons in $ZZ$ production at 
the LHC for $\sqrt{s}=13$ TeV in $4$-lepton final state. We use the cross section 
 and polarization asymmetries of the $Z$ boson  to estimate simultaneous limits on 
 anomalous coupling using Markov-Chain--Monte-Carlo (MCMC)  method  for luminosities $35.9$ fb$^{-1}$,
$150$ fb$^{-1}$, $300$ fb$^{-1}$ and $1000$ fb$^{-1}$.
The $CP$-even polarization asymmetry $A_{x^2-y^2}$ is sensitive mainly to the $CP$-odd couplings 
$f_4^{Z/\gamma}$ (quadratically) providing a probe to identify $CP$-odd nature of interaction at the LHC.  
We find that the polarization asymmetries significantly improve the estimation of anomalous couplings
should a deviation from the Standard Model (SM) be observed.
\end{abstract}

\section{Introduction}\label{sec:introduction}
The Standard Model (SM) of particle physics is a highly successful theory 
in explaining most of the phenomena in  Nature. The last milestone 
of the SM, the Higgs boson discovered~\cite{Chatrchyan:2012xdj}  at the Large 
Hadron Collider (LHC)  confirmed the Electro 
Weak Symmetry Breaking (EWSB), which is still not fully understood. 
To understand this, one needs a  precise 
measurement of the Higgs self couplings, Higgs to gauge boson couplings and gauge 
boson self couplings. Anomalous triple gauge boson couplings (aTGC)  
can play an important role in understanding the EWSB mechanism. Absence of neutral 
 aTGC at the LHC will provide support to the EWSB, while the presence of it will
 indicate new physics possibility   needed to explain many phenomena such as
 $CP$-odd, Baryogenesis, dark matter, etc. 
 The $ZZ$ and $Z\gamma$  production  are the two main
 processes where one can study the neutral aTGC.
 The neutral aTGC can be obtained by adding higher dimension effective operators to the SM~\cite{Larios:2000ni,Cata:2013sva,Degrande:2013kka}. The  neutral aTGC
start  appearing at dimension-$8$ onward  and  hence their  effect  is expected to be very small
at low energy. Alternatively, one can also parametrize the neutral aTGC  
  with dimension-$6$ (and $8$)
form factors in a model-independent way~\cite{Hagiwara:1986vm}. A Lagrangian for the parametrization discussed in 
Ref.~\cite{Hagiwara:1986vm} consisting terms up to dimension-$6$   is  given by~\cite{Gounaris:1999kf}  
\begin{eqnarray}\label{aTGC_Lagrangian}
{\cal L}_{ZZV}=
\frac{e}{M_Z^2} \Bigg [&-&
\left[f_4^\gamma \left(\partial_\mu F^{\mu \beta}\right)+f_4^Z \left(\partial_\mu Z^{\mu \beta}\right) 
\right] Z_\alpha 
\left( \partial^\alpha Z_\beta\right)
+\left[f_5^\gamma \left(\partial^\sigma F_{\sigma \mu}\right)+
f_5^Z \left(\partial^\sigma Z_{\sigma \mu}\right) \right] \wtil{Z}^{\mu \beta} Z_\beta\nonumber \\
&-&  \left[h_1^\gamma \left(\partial^\sigma F_{\sigma \mu}\right)
+h_1^Z \left(\partial^\sigma Z_{\sigma \mu}\right)\right] Z_\beta F^{\mu \beta}
-\left[h_3^\gamma  \left(\partial_\sigma F^{\sigma \rho}\right)
+ h_3^Z  \left(\partial_\sigma Z^{\sigma \rho}\right)\right] Z^\alpha
\wtil{F}_{\rho \alpha}
\Bigg ], 
\end{eqnarray} 
where $\wtil{Z}_{\mu \nu}=1/2 \epsilon_{\mu \nu \rho \sigma}Z^{\rho
	\sigma}$ ($\epsilon^{0123}=+1$)  with
$Z_{\mu\nu}=\partial_\mu Z_\nu -\partial_\nu Z_\mu$ and similarly for
the photon tensor $F_{\mu\nu}$.
Among these anomalous couplings $f_4^V$, $h_1^V$ ($V=Z,\gamma$) are $CP$-odd
in nature and $f_5^V$, $h_3^V$ are $CP$-even. The couplings
$f^V$ appear only in the $ZZ$ production process,  while $h^V$ appear in the $Z\gamma$
production process. 
The anomalous neutral triple gauge boson couplings in Eq.~(\ref{aTGC_Lagrangian}) have been widely studied in the literature
~\cite{Czyz:1988yt,Baur:1992cd,Choudhury:1994nt,Choi:1994nv,
Aihara:1995iq,Ellison:1998uy, Gounaris:1999kf,Gounaris:2000dn,Baur:2000ae,
Rizzo:1999xj,Atag:2003wm,Ananthanarayan:2004eb,Ananthanarayan:2011fr,
Ananthanarayan:2014sea,Poulose:1998sd,Senol:2013ym,Rahaman:2016pqj,Rahaman:2017qql,Ots:2006dv,
Ananthanarayan:2003wi,Chiesa:2018lcs,Chiesa:2018chc} for various colliders:  in $e^+e^-$ 
collider~\cite{Czyz:1988yt,Choudhury:1994nt,
Ananthanarayan:2004eb,Ananthanarayan:2011fr,Ananthanarayan:2014sea,Senol:2013ym,
Rahaman:2016pqj,Rahaman:2017qql,Ots:2006dv,Ananthanarayan:2003wi},  $e\gamma$ 
collider~\cite{Choi:1994nv,Rizzo:1999xj,Atag:2003wm},  $\gamma\gamma$ 
collider~\cite{Poulose:1998sd}, hadron
collider~\cite{Baur:1992cd,Ellison:1998uy,Baur:2000ae,Chiesa:2018lcs,Chiesa:2018chc} and 
both $e^+e^-$ and hadron collider~\cite{Aihara:1995iq,Gounaris:1999kf,Gounaris:2000dn}.

On the experimental side, the anomalous Lagrangian in Eq.~(\ref{aTGC_Lagrangian})
have been explored at the LEP~\cite{Acciarri:2000yu,Abbiendi:2000cu,
	Abbiendi:2003va,Achard:2004ds,Abdallah:2007ae}, the 
Tevatron~\cite{Abazov:2007ad,Aaltonen:2011zc,Abazov:2011qp} and 
the LHC~\cite{Chatrchyan:2012sga,Chatrchyan:2013nda,Aad:2013izg,
	Khachatryan:2015kea,Khachatryan:2016yro,Aaboud:2017rwm,Sirunyan:2017zjc}. 
The tightest $95~\%$ C.L. limits on anomalous couplings are obtained in  $ZZ$ production at the LHC~\cite{Sirunyan:2017zjc}  running at  
$\sqrt{s}=13$ TeV with ${\cal L}=35.9$ fb$^{-1}$ and they are given by 
\begin{align}\label{eq:CMS-limit}
− 0.0012 < f_4^Z < 0.0010,~~ − 0.0010 < f_5^Z < 0.0013,\nonumber\\
− 0.0012 < f_4^\gamma < 0.0013,~~ − 0.0012 < f_5^\gamma < 0.0013,
\end{align}
obtained by varying one parameter at a time and using only the cross section as observable.
We note that these ranges of couplings do not violate unitarity bound up to an energy scale of $10$ TeV.  
Whereas a size as large as  ${\cal O}(\pm 0.1)$ of the couplings can be  allowed if the unitarity violation 
is assumed to take place at   the energy scale of $3$ TeV, a typical energy range explored by the current $13$ TeV LHC.

The tensorial structure for some of these anomalous couplings can be generated
at higher order loop within the framework of a renormalizable theory. For example,
a fermionic triangular diagram can generate $CP$-even couplings in the SM,  some simplified
fermionic model~\cite{Corbett:2017ecn}, the Minimal Supersymmetric SM 
(MSSM)~\cite{Gounaris:2000tb,Choudhury:2000bw} and Little Higgs 
model~\cite{Dutta:2009nf}. On the other hand,   $CP$-odd
couplings can be generated  at $2$ loop in the MSSM~\cite{Gounaris:2000tb}.
A $CP$-violating $ZZZ$ vertex has been studied in 2HDM in Ref.~\cite{Corbett:2017ecn,Grzadkowski:2016lpv,Belusca-Maito:2017iob}.
Besides this, the non-commutative extension of the SM
(NCSM)~\cite{Deshpande:2001mu} can also provide these anomalous coupling structures.
We note that in the EFT framework these trilinear couplings can be obtained at dimension-$8$ operator which
also contribute to quartic gauge boson  couplings $WWVV$, $ZZZ\gamma$, $ZZ\gamma\gamma$ which appear in
triple gauge boson production~\cite{Senol:2016axw,Wen:2014mha} and vector boson scattering~\cite{Perez:2018kav}, for example. 
A  complete study of these operators will require one  to include
all these processes  involving triple gauge boson couplings as well as quartic gauge boson couplings. In the effective form factor
approach as we study in this paper, however, the triple and the quartic gauge boson couplings are independent of each other
and can be studied separately.

Here we study anomalous triple gauge boson couplings in the neutral 
sector in the framework of the model independent Lagrangian in
Eq.~(\ref{aTGC_Lagrangian})  using  polarization observables of $Z$ boson~\cite{Rahaman:2016pqj,
Boudjema:2009fz,Aguilar-Saavedra:2015yza} in  $ZZ$ pair production at the LHC.
The polarization asymmetries of $Z$ and $W$ have been used earlier to study the anomalous couplings in
$ZZ/Z\gamma$ production at $e^+e^-$ collider~\cite{Rahaman:2016pqj,Rahaman:2017qql},
in $W^+W^-$ production at $e^+e^-$ collider~\cite{Abbiendi:2000ei,Rahaman:2017aab}.
The polarization asymmetries have also been used to study Higgs-gauge boson 
interaction~\cite{Nakamura:2017ihk,Rao:2018abz}, for dark matter studies~\cite{Renard:2018tae}, 
for testing the top quark mass structure~\cite{Renard:2018bsp,Renard:2018lqv}, for studies of
special interactions of massive particles~\cite{Renard:2018jxe,Renard:2018blr}, and for studies of dark matter
  and
heavy resonance~\cite{Aguilar-Saavedra:2017zkn}.
The LHC being a symmetric collider, many polarization out of $8$ polarization of $Z$ boson
cancels out, however, three of them are non zero, which are discussed in section~\ref{sect:pol-obs}.

The leading order (LO) result of the $ZZ$ pair production cross section  is way below
the result measured at the LHC~\cite{Aaboud:2017rwm,Sirunyan:2017zjc}. However, the existing 
next-to-next-to-leading  order (NNLO)~\cite{Heinrich:2017bvg,Cascioli:2014yka} results are  
comparable with the measured values at CMS~\cite{Sirunyan:2017zjc} and ATLAS~\cite{Aaboud:2017rwm}.
We, however, obtain the cross section at next-to-leading order (NLO) in the SM and in aTGC using \MGvATNLO~\cite{Alwall:2014hca}  
and have used the SM $k$-factor to match to the NNLO value.
The details of these calculations are described in section~\ref{sect:signal-background}.

The rest of the paper is organized as follows. In section~\ref{sect:pol-obs}
we give a brief overview of the polarization observables of the $Z$ boson.  In
section~\ref{sect:signal-background} we discuss LO, NLO and NNLO result for $ZZ$ production 
including aTGC and various background processes.
In section~\ref{sect:expressions-limits} we study the sensitivity of observables to couplings
and obtain one parameter as well as  simultaneous limits on the couplings. We also
study a benchmark aTGC  and investigate how polarization asymmetries can improve the estimation of  anomalous
couplings.
We conclude in section~\ref{sect:conclusion}.

\section{Polarization observables of $Z$}\label{sect:pol-obs}
The normalized production density matrix of a spin-$1$ particle (here $Z$) 
 can be written as~\cite{Bourrely:1980mr,Boudjema:2009fz}
\begin{equation}\label{eq:spin-desnity-matrix}
\rho(\lambda,\lambda^\prime)=\dfrac{1}{3}\Bigg[I_{3\times 3} +\dfrac{3}{2} \vec{p}.\vec{S}
+\sqrt{\dfrac{3}{2}} T_{ij}\big(S_iS_j+S_jS_i\big) \Bigg],
\end{equation}
where $\vec{S}=\{S_x,S_y,S_z\}$ are the spin basis, $\vec{p}=\{p_x,p_y,p_z\}$ are the vector polarizations,
 $T_{ij}$ ($2^{nd}$-rank symmetric 
traceless tensor) are tensor polarizations  and $(\lambda,\lambda^\prime)\in\{+1,0,-1\}$ are helicities of the particle.
After expansion Eq.~(\ref{eq:spin-desnity-matrix}) can be rewritten as\footnote{The choice of polarization vector is used to be
	$\epsilon_Z(\lambda=\pm1)=\frac{1}{\sqrt{2}}\{0,\mp 1,-i,0\}$.}
\begin{eqnarray}
\label{Polarization_matrix}
\rho(\lambda,\lambda') =
\renewcommand{\arraystretch}{1.5}
 \left[
\begin{tabular}{lll}
$\frac{1}{3}+\frac{p_z}{2}+\frac{T_{zz}}{\sqrt{6}}$ &
$\frac{p_x -ip_y}{2\sqrt{2}}+\frac{T_{xz}-iT_{yz}}{\sqrt{3}}$ &
$\frac{T_{xx}-T_{yy}-2iT_{xy}}{\sqrt{6}}$ \\
$\frac{p_x +ip_y}{2\sqrt{2}}+\frac{T_{xz}+iT_{yz}}{\sqrt{3}}$ &
$\frac{1}{3}-\frac{2 T_{zz}}{\sqrt{6}}$ &
$\frac{p_x -ip_y}{2\sqrt{2}}-\frac{T_{xz}-iT_{yz}}{\sqrt{3}}$ \\
$\frac{T_{xx}-T_{yy}+2iT_{xy}}{\sqrt{6}}$ &
$\frac{p_x +ip_y}{2\sqrt{2}}-\frac{T_{xz}+iT_{yz}}{\sqrt{3}}$ &
$\frac{1}{3}-\frac{p_z}{2}+\frac{T_{zz}}{\sqrt{6}}$
\end{tabular}\right].
\end{eqnarray}
The Eq.~(\ref{Polarization_matrix}) is called a polarization density matrix of a spin-$1$ particle.
The $8$ independent polarizations $p_x,~p_y,~p_z$ and $T_{xy}$, $T_{xz}$, $T_{yz}$, 
$T_{xx}-T_{yy}$ and $T_{zz }$ can be calculated from a production density matrix
of the particle in any production process~\cite{Rahaman:2017qql,Aguilar-Saavedra:2015yza}.
The laboratory (Lab) frame and centre-of-mass frame (CM) being different at the LHC, the polarization
calculated at CM frame will not be the same as the polarization at Lab frame, unlike the total
cross section. The production density matrix receives a total rotation leaving the trace
invariant when boosted form CM to Lab frame.  These leads to  the polarization parameters $p_i$ and $T_{ij}$
getting transformed as~\cite{Bourrely:1980mr}
	\begin{eqnarray}
	p_i^{Lab}&=&\sum_{j} R_{ij}^Y(\omega)p_j^{CM},\nonumber\\
	T_{ij}^{Lab}&=&\sum_{k,l} R_{ik}^Y(\omega)R_{jl}^Y(\omega)T_{kl}^{CM},
	\end{eqnarray} 
	where
	\begin{eqnarray}
	\cos \omega&=& \cos\theta_{CM} \cos\theta_{Lab} +\gamma_{CM} \sin\theta_{CM} \sin\theta_{Lab},\nonumber\\
	\sin \omega&=& \frac{M}{E_{CM}}\left(\sin\theta_{CM} \cos\theta_{Lab} -\gamma_{CM} \cos\theta_{CM} \sin\theta_{Lab}\right),
	\end{eqnarray} 
 $R_{ij}^Y$ is the usual rotational matrix w.r.t. $y$-direction, $\theta$ is the polar angle of the particle  w.r.t. $z$-direction, 
 $M$ is the rest mass, $E_{CM}$ is the energy in CM frame and 
$\gamma_{CM}=1/\sqrt{1-\beta_{CM}^2}$ with $\beta_{CM}$ being boost of the CM 
frame\footnote{These properties has been used in Ref.~\cite{V.:2016wba,Velusamy:2018ksp}.}.

Combining the normalized production matrix in Eq.~(\ref{Polarization_matrix})
with normalized decay density matrix
of the  particle to a pair of fermion $f$,
the normalised  differential cross section would be~\cite{Boudjema:2009fz}
\begin{eqnarray} \label{eq:angular_distribution}
\frac{1}{\sigma} \ \frac{d\sigma}{d\Omega_f} &=&\frac{3}{8\pi} \left[
\left(\frac{2}{3}-(1-3\delta) \ \frac{T_{zz}}{\sqrt{6}}\right) + \alpha \ p_z
\cos\theta_f 
+ \sqrt{\frac{3}{2}}(1-3\delta) \ T_{zz} \cos^2\theta_f
\right.\nonumber\\
&+&\left(\alpha \ p_x + 2\sqrt{\frac{2}{3}} (1-3\delta)
\ T_{xz} \cos\theta_f\right) \sin\theta_f \ \cos\phi_f \nonumber\\
&+&\left(\alpha \ p_y + 2\sqrt{\frac{2}{3}} (1-3\delta)
\ T_{yz} \cos\theta_f\right) \sin\theta_f \ \sin\phi_f \nonumber\\
&+&(1-3\delta) \left(\frac{T_{xx}-T_{yy}}{\sqrt{6}} \right) \sin^2\theta_f
\cos(2\phi_f)\nonumber\\
&+&\left. \sqrt{\frac{2}{3}}(1-3\delta) \ T_{xy} \ \sin^2\theta_f \
\sin(2\phi_f) \right].
\end{eqnarray}
Here $\theta_f$, $\phi_f$ are the polar and the azimuthal orientation of the  fermion $f$,
in the rest frame of the particle ($Z$) with its would be momentum along the  $z$-direction. 
For massless final state fermions, we have  $\delta=0$ and $\alpha=(R_f^2- L_f^2)/ (R_f^2+L_f^2)$ where 
the $Zf\bar{f}$ coupling is of the type $\gamma^\mu \left(L_f \ P_L + R_f \ P_R \right)$. 
The  polarizations $p_i$ and $T_{ij}$ are calculable from  asymmetries constructed from the 
decay angular information of lepton using Eq.~(\ref{eq:angular_distribution}). For example, 
$T_{xx}-T_{yy}$ can be calculated from the asymmetry $A_{x^2-y^2}$ as 
\begin{eqnarray}\label{eq:pol_decay_Axxyy}
A_{x^2-y^2}&=&\dfrac{1}{\sigma}\Bigg[ \Bigg(\int_{-\frac{\pi}{4}}^{\frac{\pi}{4}}\dfrac{d\sigma}{d\phi}d\phi 
+ \int_{\frac{3\pi}{4}}^{\frac{5\pi}{4}}\dfrac{d\sigma}{d\phi}d\phi \Bigg)
-\Bigg( \int_{\frac{\pi}{4}}^{\frac{3\pi}{4}}\dfrac{d\sigma}{d\phi}d\phi 
+ \int_{\frac{5\pi}{4}}^{\frac{7\pi}{4}}\dfrac{d\sigma}{d\phi}d\phi \Bigg)  \Bigg] 
\nonumber\\
&\equiv & \dfrac{\sigma(\cos 2\phi>0)
-\sigma(\cos 2\phi<0)}{\sigma(\cos 2\phi>0)+\sigma(\cos 2\phi<0)}\nonumber\\
&=& \frac{1}{\pi }\sqrt{\frac{2}{3}} (1-3 \delta ) \left(T_{xx}-T_{yy}\right).
\end{eqnarray}
Likewise one can construct asymmetries corresponding to each of the polarizations
$p_i$ and $T_{ij}$, see Ref.~\cite{Rahaman:2016pqj} for details. 

The LHC being a symmetric collider, most of the polarization of $Z$ in $ZZ$ pair production
are either zero or  close to zero  except   the polarization $T_{xz}$, $T_{xx}-T_{yy}$,  and $T_{zz}$. 
To enhance the significance further we redefine the asymmetry 
corresponding to $T_{xz}$ as (see Ref.~\cite{Rahaman:2017qql}) 
\begin{eqnarray}
\wtil{A}_{xz}\equiv \frac{1}{\sigma}\bigg(\sigma(c_{\theta_Z}\times c_{\theta_f}c_{\phi_f}> 0)
-\sigma(c_{\theta_Z} \times c_{\theta_f}c_{\phi_f}< 0)\bigg),\hspace{1cm}
\end{eqnarray}
where $c_{\theta_Z}$ is the cosine of $Z$ boson polar angle in the lab frame.
To get the momentum direction of $Z$ boson,
one needs a reference axis ($z$-axis), but we can not assign a direction at the LHC because it is a
symmetric collider. So we consider the direction of boost   of the  $4l$ final state to be the proxy
for reference $z$-axis. In $q\bar{q}$ fusion, the quark is supposed to have larger momentum
then the anti-quark at the LHC, thus above proxy statistically stands for the direction of the quark
and $c_{\theta_Z}$ is measured w.r.t. the boost.

\section{Signal and background}\label{sect:signal-background}
\begin{figure}
	\centering
	\includegraphics[width=1\textwidth]{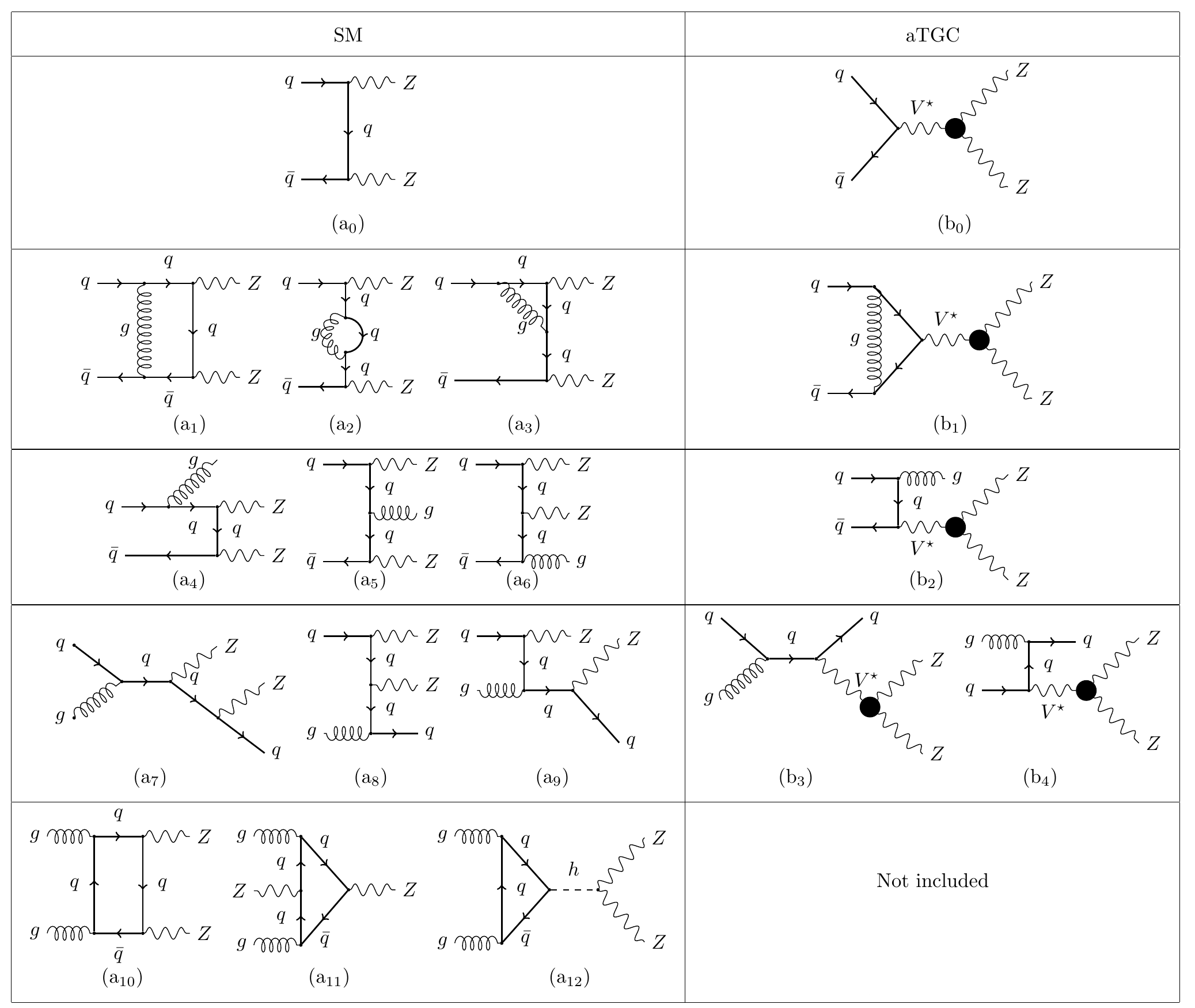}
	\caption{\label{fig:Feynman-diagram-all}
		Representative Feynman diagrams for $ZZ$ pair production at the LHC in the SM 
		($q\bar{q}$  and $gg$ initiated) as well as in aTGC ($q\bar{q}$ initiated) 
		at tree level together with NLO in QCD.
	}
\end{figure}
\begin{table}[t]\caption{\label{tab:ZZ-signal} The theoretical estimates available in   literature and experimental measurements of the
 $ZZ$ production  cross section at $\sqrt{s}=13$ TeV at the LHC. The uncertainties in the theoretical 
 estimates come from scale variation.}
\renewcommand{\arraystretch}{1.50}
\begin{tabular*}{\columnwidth}{@{\extracolsep{\fill}}llll@{}}\\ \hline
Ref. & $\sigma_{\text{LO}}$ [pb]& $\sigma_{\text{NLO}}$ [pb] & $\sigma_{\text{NNLO}}$ [pb] \\ \hline
Ref.~\cite{Heinrich:2017bvg} & $9.890_{-6.1\%}^{+4.9\%}$ & $14.51_{-2.4\%}^{+3.0\%}$ & $16.92_{-2.6\%}^{+3.2\%}$ \\ \hline
Ref.~\cite{Cascioli:2014yka}& $9.887_{-6.1\%}^{+4.9\%}$ & $14.51_{-2.4\%}^{+3.0\%}$ & $16.91_{-2.4\%}^{+3.2\%}$\\ \hline
 \end{tabular*}
\begin{tabular*}{\columnwidth}{@{\extracolsep{\fill}}llll@{}}
CMS~\cite{Sirunyan:2017zjc} &  $17.2\pm 0.5 (stat.) \pm 0.7 (syst.) 
	\pm 0.4 (lumi.) $ & \\ \hline
ATLAS~\cite{Aaboud:2017rwm} & $17.3 \pm 0.6(stat.) \pm 0.5(syst.) \pm 0.6(lumi.)$& \\ \hline
\end{tabular*}
\end{table}
We are interested in studying anomalous triple gauge boson couplings in $ZZ$ pair
production at the LHC. The tree level standard model contribution to this process comes from the representative
diagram (a$_0$) in Fig.~\ref{fig:Feynman-diagram-all}, while the tree level aTGC contribution is shown in the diagram (b$_0$). 
Needless to say, the tree level cross section in the SM is way below the
measured cross section at the LHC, because  QCD corrections are very high in this process.
In the SM, at NLO (${\cal O}(\alpha_s)$), virtual  contributions come from the representative diagrams (a$_{1}$--a$_{3}$) 
and real contributions come  from (a$_{4}$--a$_{9}$) in the    $q\bar{q}$ initiated sub-process.
The  $gg$ initiated sub-process  appears
at $1$-loop level,   the diagrams (a$_{10}$--a$_{12}$), and contributes at ${\cal O}(\alpha_s^2)$. 
The LO, NLO and NNLO results from theoretical calculation  available in   literature~\cite{Heinrich:2017bvg,Cascioli:2014yka} 
for $ZZ$ production cross section  at $\sqrt{s}=13$ TeV for a  $pp$ collider are listed
in Table~\ref{tab:ZZ-signal}. The recent experimental
measurement from CMS~\cite{Sirunyan:2017zjc}
and ATLAS~\cite{Aaboud:2017rwm} are also shown for comparison.
The cross section at
NLO  receives as much as $\sim 46~\%$ correction over LO and further the NNLO cross section
receives  $\sim 16~\%$  correction over the NLO result. At NNLO the $q\bar{q}$ sub-process receives
$10~\%$ correction~\cite{Cascioli:2014yka} over NLO and the $gg$ initiated ${\cal O}(\alpha_s^3)$ sub-process receives $70~\%$
correction~\cite{Caola:2015psa} over it's ${\cal O}(\alpha_s^2)$ result.
The LO and NLO results obtained in \MGvATNLO~v2.6.2 with pdf (parton-distribution-function) sets NNPDF23 are 
\begin{eqnarray}\label{eq:result-mg5-zz}
\sigma_{{\cal O}(\alpha_{s}^0)}^{q\bar{q}\to ZZ} &=& 9.341_{-5.3\%}^{+4.3\%}  ~~\text{pb},\nonumber\\  
\sigma_{{\cal O}(\alpha_s)}^{q\bar{q}\to ZZ} &=& 13.65_{-3.6\%}^{+3.2\%}  ~~\text{pb}, \nonumber\\ 
\sigma_{{\cal O}(\alpha_s^2)}^{gg\to ZZ} &=& 1.142_{-18.7\%}^{+24.5\% } ~~\text{pb},\nonumber\\   
\sigma_{mixed_1}^{q\bar{q}+gg\to ZZ}&=&
\sigma_{{\cal O}(\alpha_s)}^{q\bar{q}\to ZZ}+ \sigma_{{\cal O}(\alpha_s^2)}^{gg\to ZZ}\nonumber\\
 &=&14.79_{-4.7\%}^{+4.8\%} ~~\text{pb}. 
\end{eqnarray}
The  errors in the subscript and superscript on the  cross section are the uncertainty 
from scale variation. 
The total cross section combining the $q\bar{q}$  sub-process at ${\cal O}(\alpha_s^2)$
 with $gg$ at ${\cal O}(\alpha_s^3)$  is given by
\begin{eqnarray}
\sigma_{mixed_2}^{q\bar{q}+gg\to ZZ}&=&
\underbrace{\sigma_{{\cal O}(\alpha_s)}^{q\bar{q}\to ZZ}\times 1.1}_{{\cal O}(\alpha_s^2)} \ \ + \ \ \underbrace{ \sigma_{{\cal O}(\alpha_s^2)}^{gg\to ZZ}\times 1.7}_{{\cal O}(\alpha_s^3)} \nonumber\\
&=&16.96_{-5.3\%}^{+5.6\%} ~~\text{pb}. 
\end{eqnarray}
The  aTGC  has also a substantial NLO QCD correction and they come from 
the diagram (b$_{2}$) at 1 loop level and from   (b$_{2}$--b$_{4}$) as the real radiative process. 
The aTGC effect is not included in the $gg$ process where the aTGC may come from a similar 
diagram with   $h\to ZZ$ in Fig.~\ref{fig:Feynman-diagram-all}(a$_{12}$) but $h$ replaced
with a $Z$. As an example of NLO QCD correction of aTGC in this process, we obtain cross section
at $\sqrt{s}=13$ TeV  with all couplings $f_i^V=0.001$. The  cross section for only aTGC part, $(\sigma^{\text{aTGC}}-\sigma^{\text{SM}})$
at LO and NLO are  $71.82$~fb $(0.77~\%)$ and $99.94$~fb $(0.73~\%)$, respectively. Thus  NLO 
result comes with a substantial amount ($\sim 39~\%$) of QCD correction over LO at this given aTGC point.

The signal  consists of $4l$ ($2e2\mu/4e/4\mu$) final state which includes  
 $ZZ$,  $Z\gamma^\star$, and $\gamma^\star \gamma^\star$ processes. The signal events are generated  
 in \MGvATNLO\\with pdf sets NNPDF23 in the SM as well as in the aTGC as $pp\to VV \to 2e2\mu$  
 ($V=Z/\gamma^\star$) at NLO in QCD in $q\bar{q}$, $qg$ as well as in $1$-loop $gg$ initiated process
with the following basic cuts (in accordance with Ref.~\cite{Sirunyan:2017zjc}),
 \begin{itemize}
 	\item $p_T^l>10$  GeV,
 	hardest  $p_T^l>20$  GeV,
 	and second hardest  $p_T^l>12$  GeV,
 	\item $|\eta_e|<2.5$, $|\eta_{\mu}|<2.4$,
 	\item $\Delta R (e,\mu)>0.05$, $\Delta R (l^+,l^-)>0.02$.
 \end{itemize}
To select the $ZZ$ final state from the above generated signal we further put a constraint on invariant
mass of same flavoured oppositely charged leptons pair with
 \begin{itemize}
\item $60$ GeV $< M_{l^+l^-}< 120$ GeV. 
\end{itemize}
The  $2e2\mu$ cross section up to a factor of two is used as the proxy for the $4l$ cross section
for the ease of event generation and related handling.

The background event consisting $t\bar{t}Z$ and $WWZ$ with leptonic decay 
are generated at LO in \MGvATNLO~   with NNPDF23 with the same sets of 
cuts as applied to the signal and their cross section is matched to NLO in QCD with a $k$-factor of $1.4$. 
This  $k$-factor estimation was done at the production level.
 We have estimated the total cross section of the signal in the SM to be
\begin{eqnarray}
\sigma(pp\to ZZ\to 4l)_{{\cal O}(\alpha_s)}^{q\bar{q}} &=& 28.39~~\text{fb}, \nonumber \\
\sigma(pp\to ZZ\to 4l)_{{\cal O}(\alpha_s^2)}^{gg} &=&  1.452~~\text{fb}, \nonumber \\
\sigma(pp\to ZZ\to 4l)_{mixed_1}^{q\bar{q}+gg} &=&  29.85~~\text{fb}, \nonumber \\
\sigma(pp\to ZZ\to 4l)_{mixed_2}^{q\bar{q}+gg} &=&  33.70~~\text{fb}.
\end{eqnarray}
  The background cross section  at NLO  is estimated to be 
 \begin{align}
 \sigma(pp\to t\bar{t}Z+WWZ\to 4l+ \cancel{\it{E}}_{T})_{NLO}= 0.020~~\text{fb}. 
 \end{align}
 The values of various parameters  used   for the generation of signal and background are
 \begin{itemize}
 	\item $M_Z=91.1876$ GeV, $M_H=125.0$ GeV,
 	\item $G_F=1.16639\times 10^{-5}$ GeV$^{-2}$, $\alpha_{em}=1/132.507$,\\ $\alpha_s=0.118$,
 	\item $\Gamma_Z=2.441404$ GeV, $\Gamma_H=6.382339$ MeV.
 \end{itemize}
The renormalization and factorization scale is set to $\sum M_i^T/2$, $M_i^T$ are the transverse mass  of all final state particles.

In our  analysis, the total cross section in the SM including the aTGC is taken 
as\footnote{$mixed_1 \approx q\bar{q}({\cal O}(\alpha_s)) + gg({\cal O}(\alpha_s^2))$, 
$mixed_2 \approx q\bar{q}({\cal O}(\alpha_s^2)) + gg({\cal O}(\alpha_s^3))$}
\begin{align}\label{eq:sigma-setup}
\sigma_{\text{Tot}}=\sigma^{\text{SM}}_{mixed_2} + (\sigma_{\text{NLO}}^{\text{aTGC}} 
- \sigma_{\text{NLO}}^{\text{SM}}),
\end{align}
 the SM is considered at order $mixed_2$, whereas the aTGC contribution
along with its interference with the SM are considered at NLO in QCD (as the NNLO contribution is 
not known with aTGC).

We will use polarization asymmetries as described in the previous section in our
analysis. Assuming that the NNLO effect cancels away because of the ratio
of two cross section, we will use the asymmetries  as
\begin{align}\label{eq:asymmetry-setup}
A_i=\frac{\Delta\sigma_i^{mixed_1}}{\sigma^{mixed_1}}.
\end{align}
We use total cross section at $mixed_2$ order   and asymmetries 
at $mixed_1$ order to put constrain  on the anomalous couplings. We note that the
$Z$ boson momenta is  required to  be reconstructed to obtain its polarization asymmetries, which require the
right pairing of two oppositely charged leptons coming from a same $Z$ boson in $4e/4\mu$  channel. The right paring of
leptons for the $Z$ boson in the same flavoured channel is possible with $\sim 95.5~\%$ for $M_{4l}>300$ GeV and
$\sim 99~\%$ for $M_{4l}>700$ GeV for both SM and aTGC by requiring a smaller value of $|M_Z-M_{l^+l^-}|$. This small miss pairing is neglected as it allows to use  the $2e2\mu$ channel as a proxy for the full $4l$ final state with good enough accuracy.

\subsection{Effect of aTGC in distributions}
\begin{figure}
	\centering
	\includegraphics[width=0.48\textwidth]{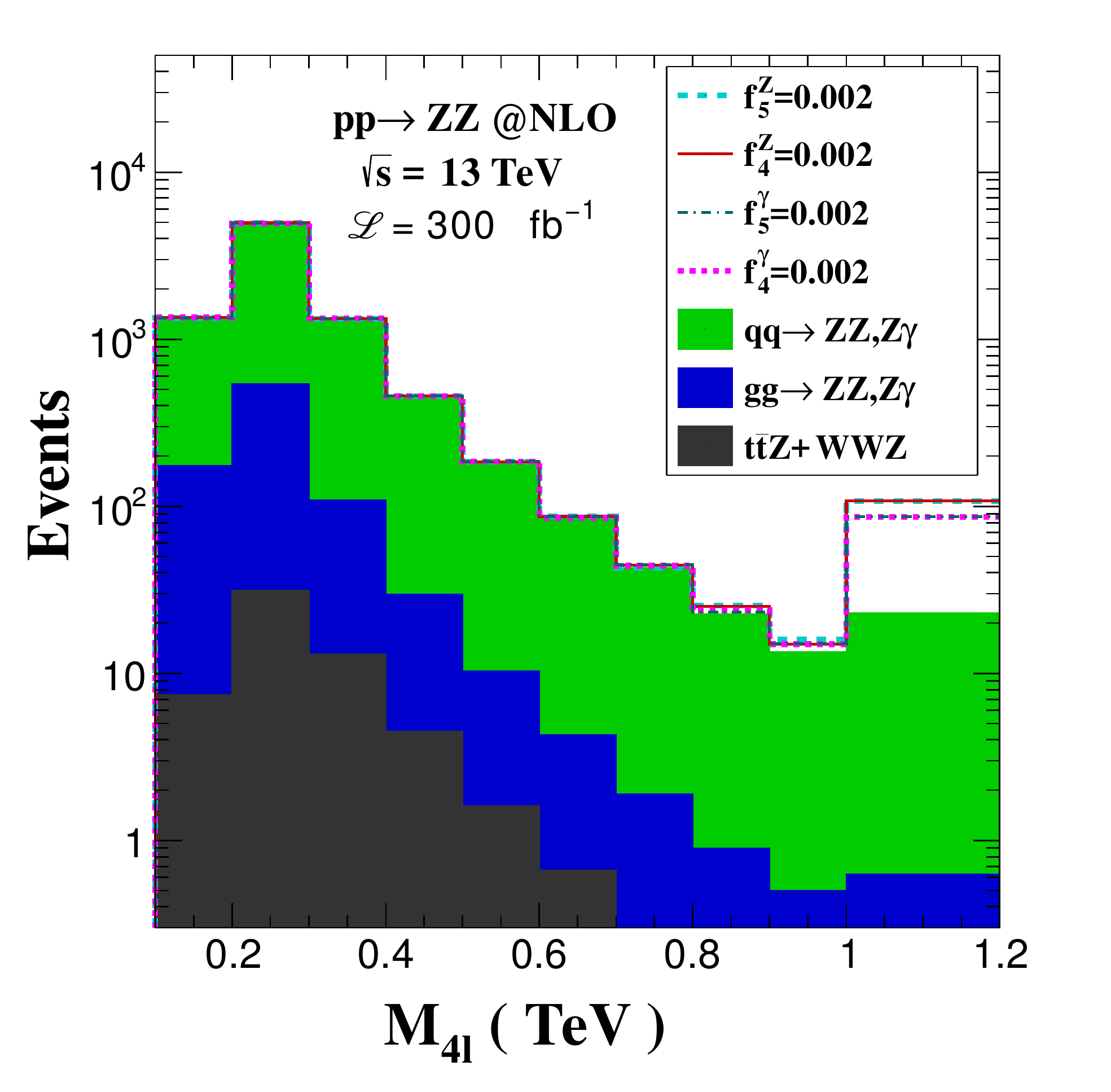}	
	\includegraphics[width=0.48\textwidth]{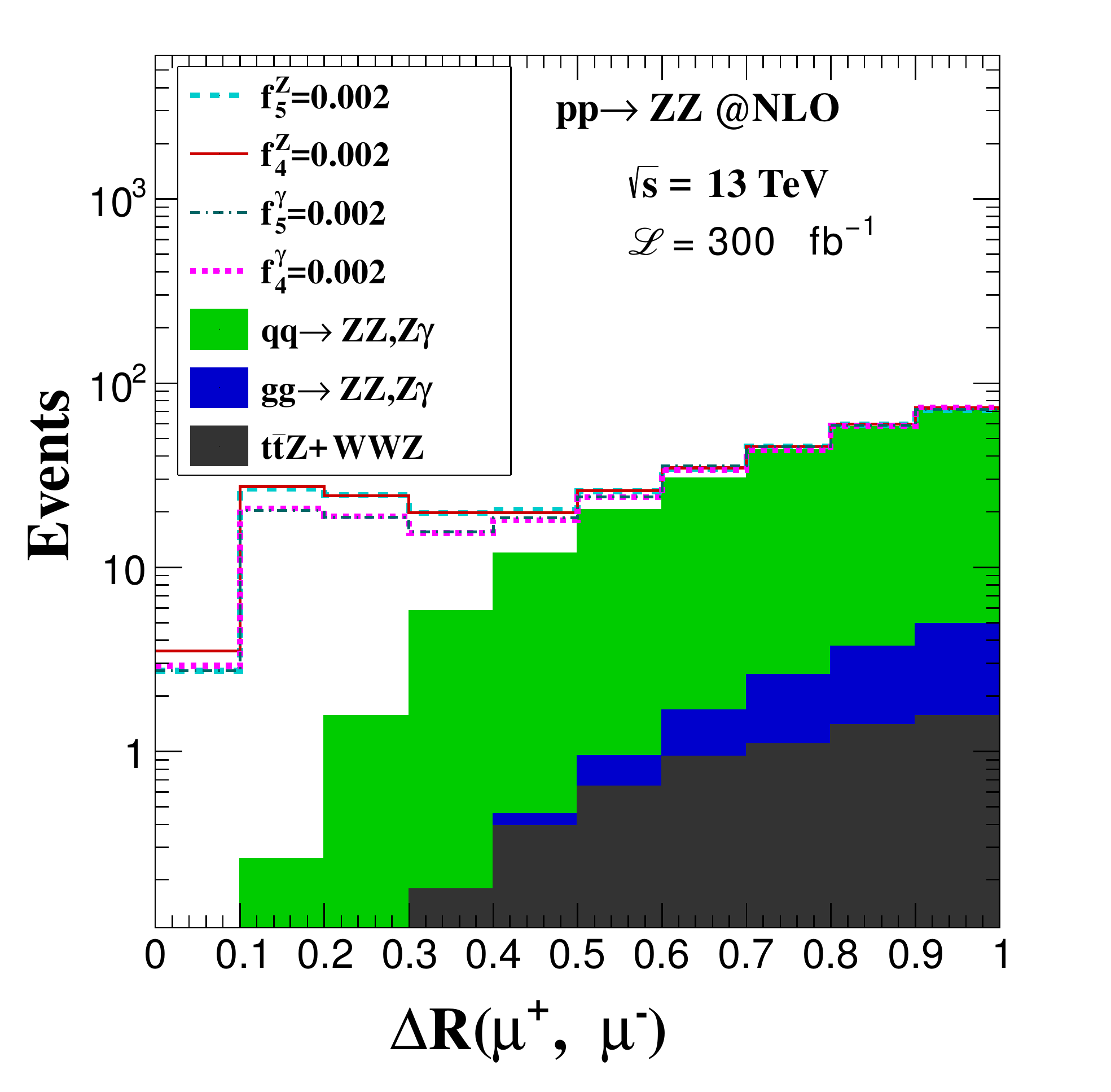}
	\caption{\label{fig:m4lDrll}
		$M_{4l}$ ({\em left-panel}) and $\Delta R$ between $\mu^+,\mu^-$ ({\em right-panel}) distribution in
		$ZZ$ production at the LHC at $\sqrt{s}=13$ TeV and  
		${\cal L}=300$ fb$^{-1}$ at NLO in QCD. The SM signal and background are shown in shaded region,
		while  aTGC contributions are shown with different line types. }
\end{figure}
\begin{figure}
	\centering
	\includegraphics[width=0.495\textwidth]{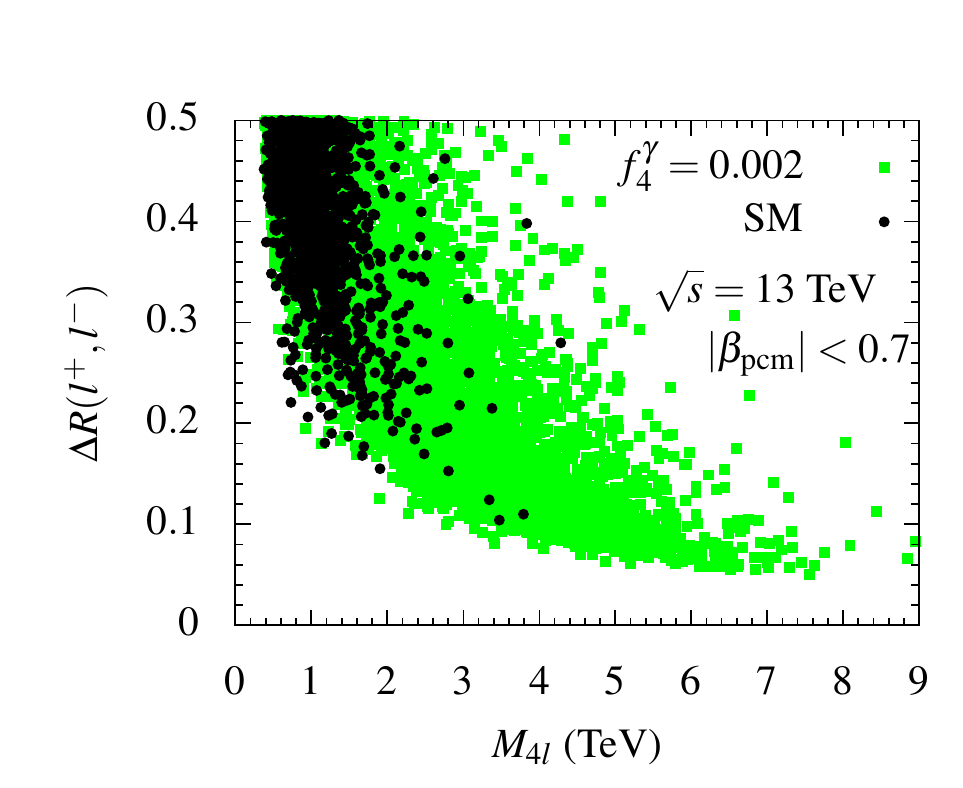}
	\includegraphics[width=0.49\textwidth]{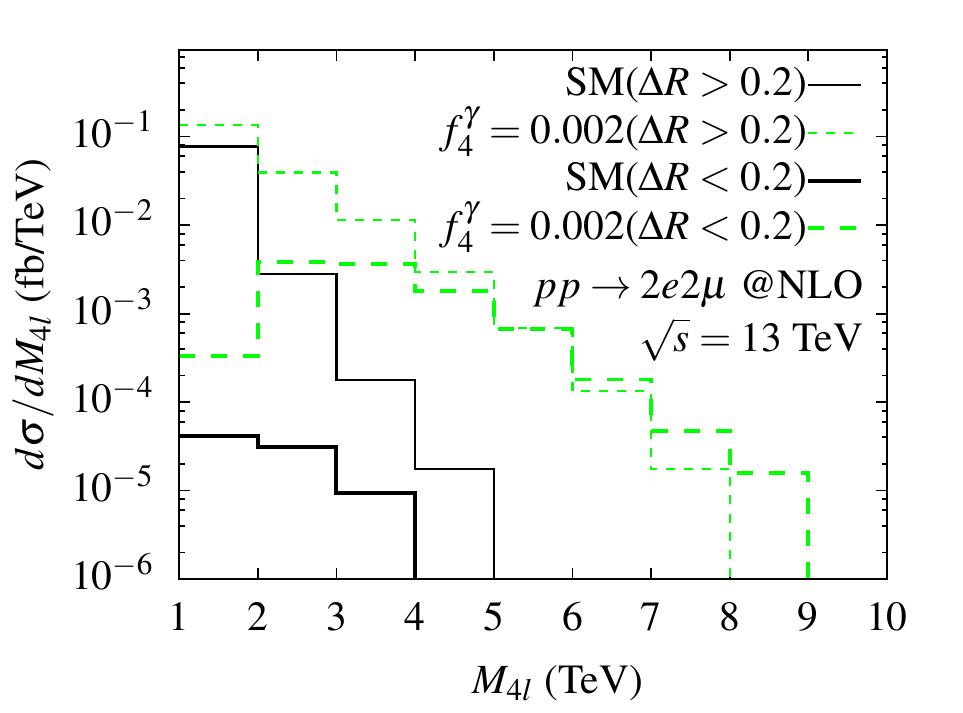}
	\caption{\label{fig:m4lvsDrr}
$M_{4l}$ vs $\Delta R$  scattered plot ({\em left}) and $M_{4l}$ distribution for $\Delta R(l^+,l^-) \gtrless 0.2$ ({\em right}) in
$ZZ$ production at the LHC at $\sqrt{s}=13$ TeV for the SM and for aTGC with $f_4^\gamma=0.002$.}
\end{figure}
The effect of aTGC on observables varies with energy scale. 
We study the effect of aTGC on various observables in their distribution and determine the signal region.
In Fig.~\ref{fig:m4lDrll}  
we show four lepton invariant mass ($M_{4l}$) or  centre-of-mass energy ($\sqrt{\hat{s}}$)
distribution ({\em left-panel})  and $\Delta R$ distribution of $\mu^+\mu^-$ pair ({\em right-panel})   
at $\sqrt{s}=13$ TeV  for the  SM along with background 
$t\bar{t}Z+WWZ$ and some benchmark aTGC points for events 
normalized to luminosity  $300$ fb$^{-1}$ using {\tt \sc{MadAnalysis5}}~\cite{Conte:2012fm}.  
The   $gg$ contribution  is at its LO (${\cal O}(\alpha_s^2)$), while all other contributions are 
shown at NLO (${\cal O}(\alpha_s)$).
The $q\bar{q} \to ZZ,~Z\gamma$ contribution is shown in {\em green band}, $gg \to ZZ,Z\gamma$ is 
in {\em blue band} and the background $t\bar{t}Z+WWZ$ contribution is shown in {\em grey band}. 
 The aTGC contribution for various choices are   shown in {\em dashed}/cyan ($f_5^Z=0.002$),   {\em solid}/red ($f_4^Z=0.002$),
  {\em dashed-dotted}/dark-green ($f_5^\gamma=0.002$)  and 
 {\em small-dashed}/magenta ($f_4^\gamma=0.002$).   
 For the $M_{4l}$ distribution in left, all events above $1$ TeV are added  in the last bin.
 All the aTGC benchmark i.e., $f_i^V=0.002$ are not visibly different 
than the SM $q\bar{q}$ contribution upto $\sqrt{\hat{s}}=0.8$ TeV and there are significant 
excess of events in the last bin, i.e., above $\sqrt{\hat{s}}=1$ TeV. 
This is due to momentum dependence~\cite{Rahaman:2016pqj} of the interaction vertex
that leads to increasing contribution at higher momentum transfer.
 In the distribution of $\Delta R (\mu^+,\mu^-)$ in the {\em right-panel},
 the effect of aTGC is higher for lower $\Delta R$  (below $0.5$). 
 In the $ZZ$ process,  the $Z$ bosons are highly boosted for larger $\sqrt{\hat{s}}$
 and their decay products are collimated leading to a smaller $\Delta R$ separation between
 the decay leptons. To see this kinematic effect we plot events in $M_{4l}$ - $\Delta R$
 plane in Fig.~\ref{fig:m4lvsDrr} ({\em left-panel}). Here,  we choose a minimum $\Delta R$ between $e$ pair and $\mu$ pair
 event by event. We note that additional events coming from aTGC contributions have higher 
 $M_{4l}$ and lower $\Delta R$ between leptons. For $\Delta R<0.2$ most of the events
 contribute to the $M_{4l}>1$ TeV bin and they are dominantly coming from aTGC,  Fig.~\ref{fig:m4lvsDrr} ({\em right-panel}). Thus
 we can choose $M_{4l}>1$ TeV to be the signal region.

\section{Sensitivity  of observables and limits on the anomalous couplings}\label{sect:expressions-limits}
\begin{figure}
\centering
\includegraphics[width=0.45\textwidth]{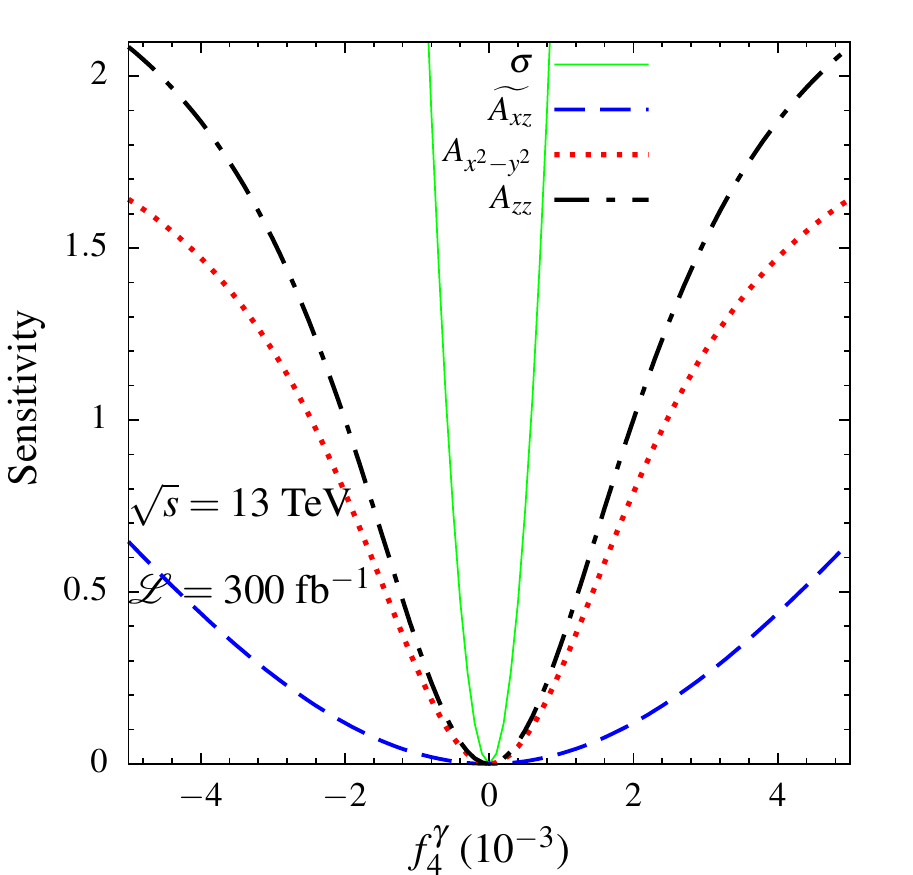}
\includegraphics[width=0.45\textwidth]{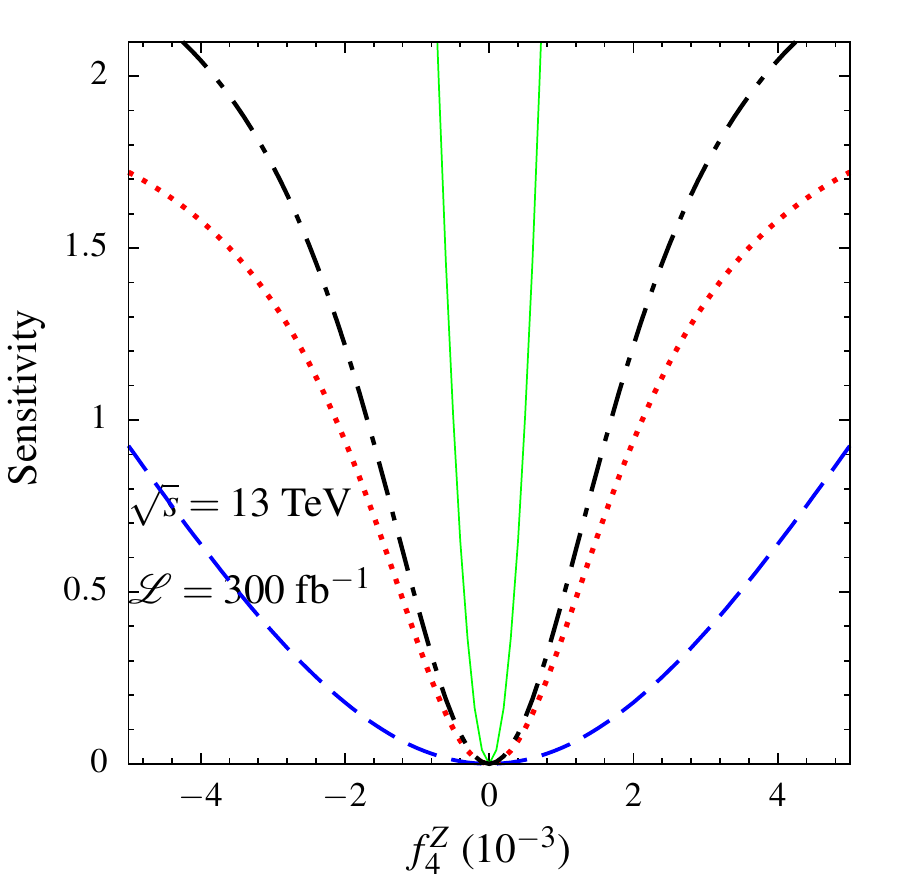}
\includegraphics[width=0.45\textwidth]{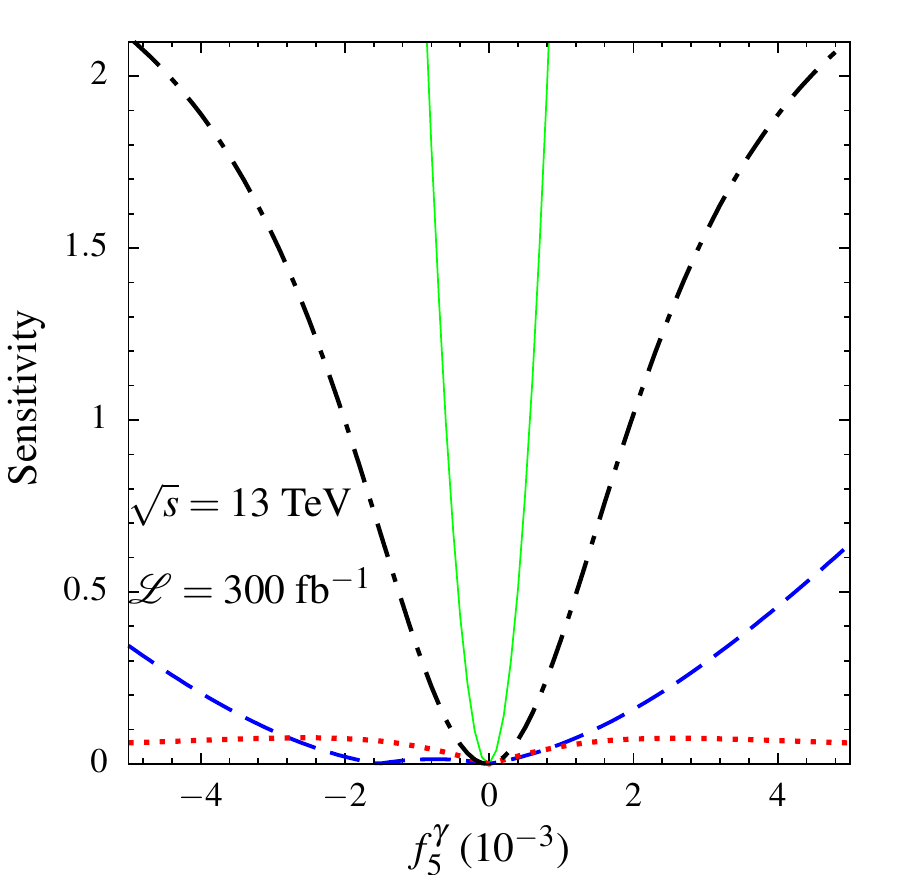}
\includegraphics[width=0.45\textwidth]{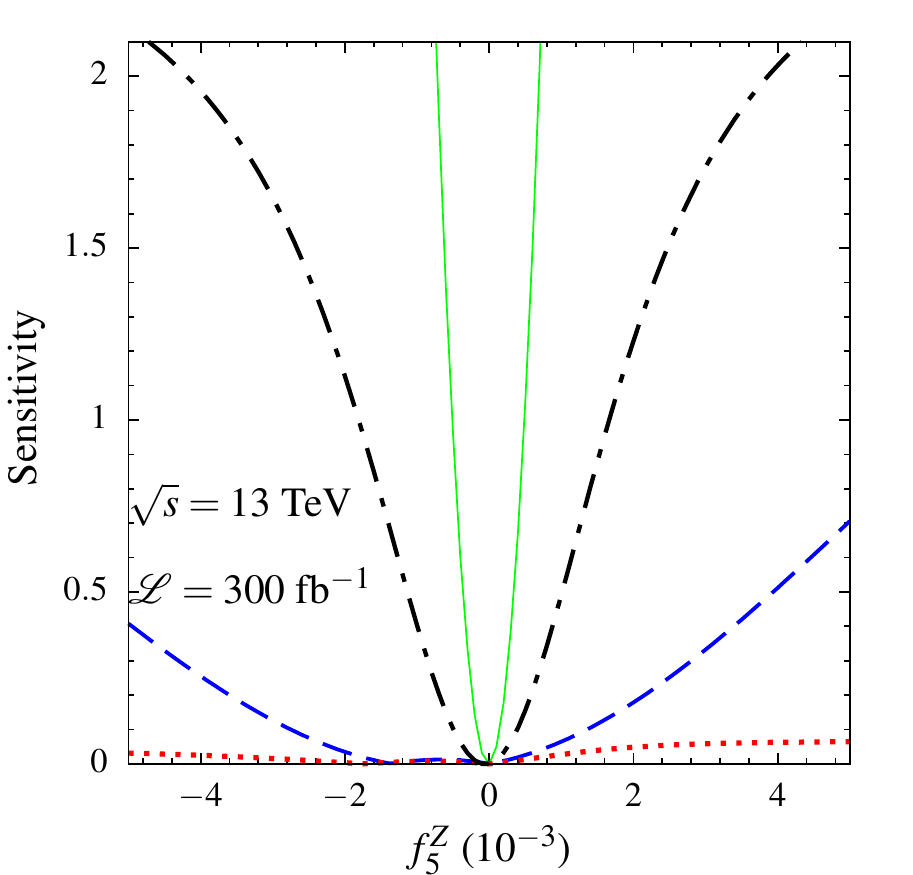}
\caption{\label{fig:sensitivity-L300}The  sensitivity  of the cross section and the polarization observables
 to the anomalous  couplings at $\sqrt{s}=13$ TeV and  ${\cal L}=300$ fb$^{-1}$ in $ZZ$ production at the LHC. }
\end{figure}
In this analysis, the set of observables consist of the cross section and polarization 
asymmetries $\wtil{A_{xz}}$, $A_{x^2-y^2}$, and $A_{zz}$.
The signal region for the cross section $\sigma$ is chosen to be $M_{4l} >1$ TeV as 
we have discussed in the previous 
section. In case of asymmetries, we choose the signal region as $M_{4l} >0.3$ TeV for $\wtil{A_{xz}}$ and 
$M_{4l} >0.7$ TeV for $A_{x^2-y^2}$ and $A_{zz}$ as the effect of aTGC is found to be best
in these region corresponding to these asymmetries. 
The expression for the cross section and the polarization asymmetries as a function of couplings are obtained by
numerical fitting the data generated by \MGvATNLO. The events are generated   for
different set of values of  the couplings $f_i^V=(f_4^\gamma,~f_4^Z,~f_5^\gamma,~f_5^Z)$ and then
various cross sections, i.e., the total cross section and the numerator of the asymmetries, ${\cal O} $, are fitted as
\begin{align}\label{eq:gen-fit}
{\cal O} = {\cal O}_0 + f_i^V \times {\cal O}_i + f_i^V\times f_j^V \times {\cal O}_{ij},     
\end{align}
in general, where ${\cal O}_0$ is the value of corresponding cross sections in the SM.
 The observables, considered here, are all $CP$-even in nature which leads to the modification of Eq.~(\ref{eq:gen-fit})
 as 
 \begin{align}\label{eq:cp-even-fit}
 {\cal O} = {\cal O}_0 + f_5^V \times {\cal O}_5^V + f_4^\gamma  f_4^Z\times {\cal O}_4^{\gamma,Z}
 + f_5^\gamma  f_5^Z\times {\cal O}_5^{\gamma,Z} + (f_i^V)^2 \times {\cal O}_i^{VV},
 \end{align}
as the $f_4^V$ are $CP$-odd, while $f_5^V$ are $CP$-even couplings reducing the unknown from $15$ 
to $9$ to be solved.
The numerical expressions of the cross section and the asymmetries as a function of the couplings
are  given in~\ref{app:a}.
The observables are obtained up to   ${\cal O}(\Lambda^{-4})$, i.e., quadratic in dimension-$6$. 
In practice, one should consider the effect of dimension-$8$ contribution at linear order. However, we choose to work
with only dimension-$6$  in couplings with a contribution up to quadratic so as to compare the results with the  current LHC constraints on
dimension-$6$ parameters~\cite{Sirunyan:2017zjc}. 
A note on keeping terms up to quadratic in  couplings, and not terminating at linear order, is presented in~\ref{app:b}.

\subsection{Sensitivity of observables to the  couplings}
The sensitivity of an observable
${\cal O}(f_i)$ to coupling $f_i$ is defined as
\begin{align}
{\cal S}{\cal O}(f_i)=\dfrac{|{\cal O}(f_i)-{\cal O}(f_i=0)|}{\delta {\cal O}},
\end{align}
where $\delta {\cal O}$ is the estimated error in ${\cal O}$. For cross 
section and  asymmetries, the errors are
\begin{align}
\delta\sigma=\sqrt{\dfrac{\sigma}{{\cal L}} + (\epsilon_\sigma\sigma)^2}~~~\text{and}~~~
\delta A_i=\sqrt{\dfrac{1-A_i^2}{{\cal L}\times \sigma}+\epsilon_A^2},
\end{align}
where ${\cal L}$ is the integrated luminosity and $\epsilon_\sigma$ and $\epsilon_A$ are the systematic
uncertainty for the cross section and the asymmetries, respectively.
 We consider  $\epsilon_\sigma=5~\%$~\cite{Sirunyan:2017zjc}  and 
$\epsilon_A=2~\%$ in this analysis  as a benchmark.  The sensitivity of all the observables to the 
couplings are shown in Fig.~\ref{fig:sensitivity-L300} for ${\cal L}=300$ fb$^{-1}$.
We find asymmetries to be less sensitive than the cross section to the couplings
 and thus cross section wins in putting limits on the couplings. 
 The sensitivity curve of all the couplings in each observable
 are symmetric about zero as $f_4^V$ (being $CP$-odd) does not appear in linear in any observables and also  the
 linear contribution from $f_5^V$ are negligibly small compared to their quadratic contribution (see \ref{app:a}). 
For example, the coefficient of $f_5^V$ are $\sim 1$ in $\sigma (M_{4l} > 1~\text{TeV})$ (Eq.~(\ref{eq:sigma-1TeV})), while the
coefficient of $(f_5^V)^2$ are $\sim 5\times 10^{4}$. Thus even at 
 $f_5^V=10^{-3}$ the quadratic contribution is $50$ times stronger than the linear one.
 Although the asymmetries are not strongly sensitive to the couplings as the cross section, they are useful in 
the measurement of the anomalous couplings, which will be discussed in the next section.
 
  It is noteworthy to mention that the sensitivity
 of $A_{x^2-y^2}$ are flat and negligible for $CP$-even couplings $f_5^V$, while they  vary   significantly
  for $CP$-odd couplings $f_4^V$. Thus the asymmetry 
 $A_{x^2-y^2}$, although a $CP$-even observables, is able to distinguish between $CP$-odd and $CP$-even interactions
 in the $ZZ$ production at the LHC.
  \begin{table}\caption{\label{tab:single-limits} One parameter limits ($10^{-3}$) at $95~\%$ C.L. on anomalous couplings
  		in $ZZ$ production at the LHC at $\sqrt{s}=13$ TeV for various luminosities.}
  	\renewcommand{\arraystretch}{1.50}
  	\begin{tabular*}{\columnwidth}{@{\extracolsep{\fill}}lllll@{}}\hline
  		param / ${\cal L}$ & $35.9$ fb$^{-1}$ & $150$ fb$^{-1}$ & $300$ fb$^{-1}$ & $1000$ fb$^{-1}$\\\hline
  		$f_4^\gamma$ &$ _{-1.20}^{+1.22}$&$ _{-0.85}^{+0.85}$&$ _{-0.72 }^{+ 0.72}$&$ _{-0.55 }^{+ 0.55}$	 \\\hline
  		$f_5^\gamma$ &$ _{-1.23}^{+1.21}$&$ _{-0.87}^{+0.84}$&$ _{-0.74 }^{+ 0.71}$&$ _{-0.57 }^{+ 0.54}$	 \\\hline
  		$f_4^Z$      &$ _{-1.03}^{+1.04}$&$ _{-0.72}^{+0.73}$&$ _{-0.61 }^{+ 0.62}$&$ _{-0.47 }^{+ 0.47}$	 \\\hline
  		$f_5^Z$      &$ _{-1.05}^{+1.03}$&$ _{-0.74}^{+0.72}$&$ _{-0.63 }^{+ 0.61}$&$ _{-0.49 }^{+ 0.46}$	 \\\hline
  	\end{tabular*}
  \end{table}

We use the total $\chi^2$ as
\begin{equation}
\chi^2(f_i)=\sum_{j} \left[ {\cal S}{\cal O}_j(f_i) \right]^2
\end{equation}
to obtain the single parameter limits on the couplings
  by varying one parameter at a time and keeping all other  to their SM values.
 The single parameter limits thus  obtained on all the anomalous couplings at $95~\%$ C.L.  
  for four benchmark luminosities ${\cal L}=35.9$ fb$^{-1}$,  
 $150$ fb$^{-1}$, $300$ fb$^{-1}$ and $1000$ fb$^{-1}$ are presented in Table~\ref{tab:single-limits}. 
 The  limit at ${\cal L}=35.9$ fb$^{-1}$ given in the first column of Table~\ref{tab:single-limits}
 are comparable to the tightest limit available at the LHC  by CMS~\cite{Heinrich:2017bvg} given in
 Eq.~(\ref{eq:CMS-limit}).  
\subsection{Limits on the couplings from MCMC}
\begin{figure*}
	\centering
	\includegraphics[width=0.48\textwidth]{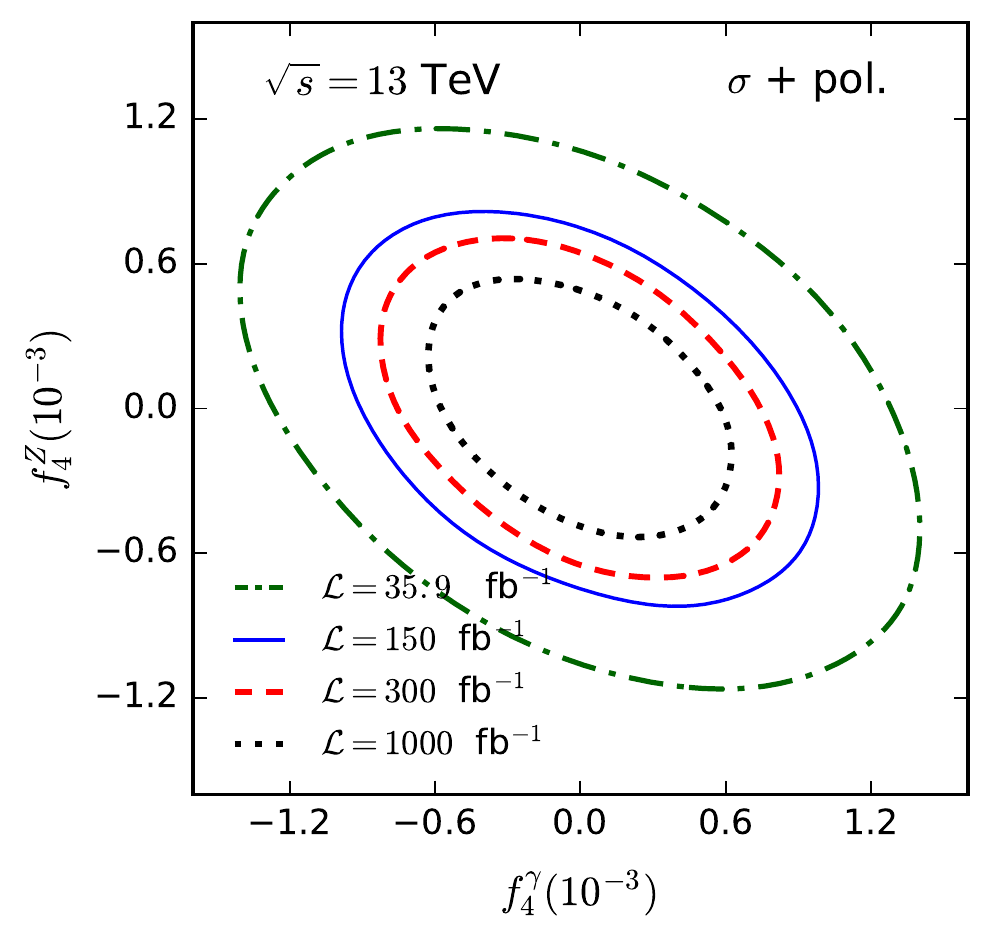}	
	\includegraphics[width=0.48\textwidth]{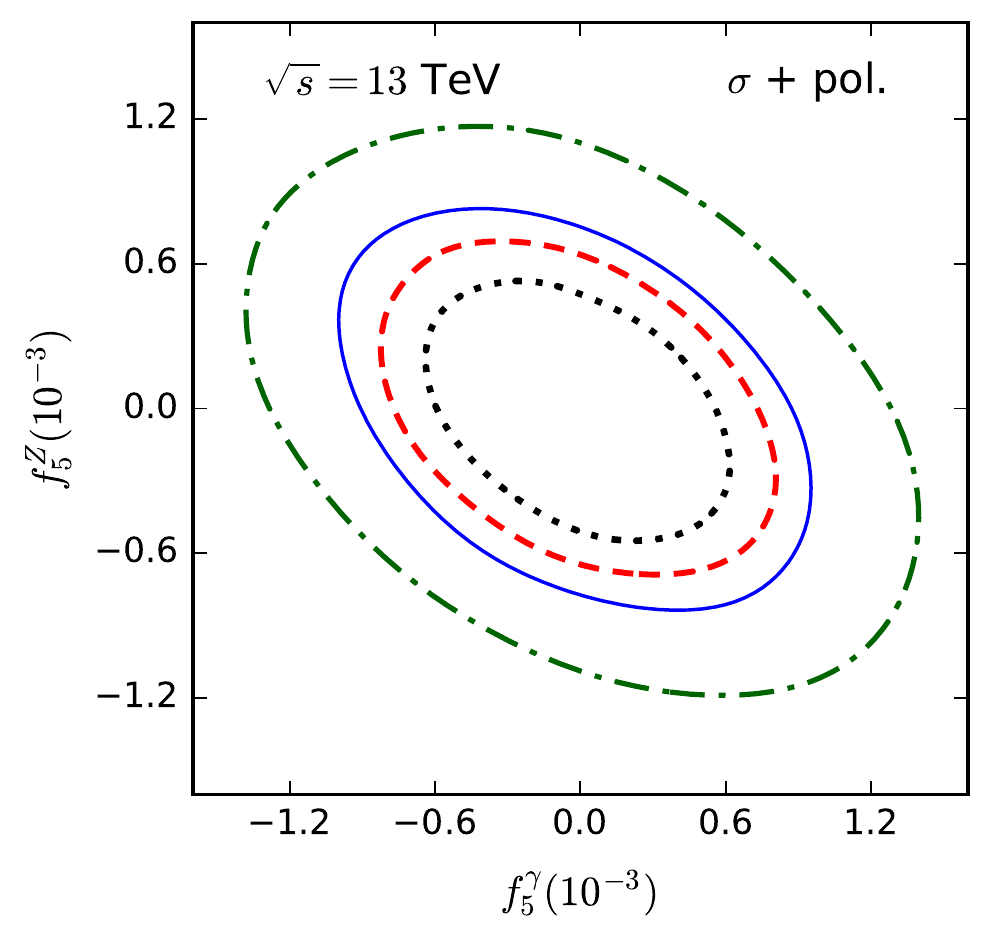}
	\caption{\label{fig:2dmcmcplot-sigwithpol} Two dimensional marginalised contours at
		$95~\%$ BCI from MCMC using the cross section $\sigma$ along with polarization asymmetries (pol.)  
		at $\sqrt{s}=13$ TeV for various luminosities in $ZZ$ production at the LHC. }
\end{figure*}
\begin{table}\caption{\label{tab:simul-limits} Simultaneous limits ($10^{-3}$) at $95~\%$ C.L. on anomalous couplings
in $ZZ$ production at the LHC at $\sqrt{s}=13$ TeV for various luminosities from MCMC. }
\renewcommand{\arraystretch}{1.50}
\begin{tabular*}{\columnwidth}{@{\extracolsep{\fill}}lllll@{}}\hline
param / ${\cal L}$ & $35.9$ fb$^{-1}$   & $150$ fb$^{-1}$       & $300$ fb$^{-1}$       & $1000$ fb$^{-1}$\\\hline
$f_4^\gamma$      &$ _{-1.15}^{+1.17}$  &$ _{-0.81}^{+0.81}$ &$ _{-0.68 }^{+ 0.67}$&$ _{-0.52 }^{+ 0.52}$	 \\\hline
$f_5^\gamma$      &$ _{-1.13}^{+1.50}$  & $ _{-0.83}^{+0.78}$&$ _{-0.68 }^{+ 0.66}$&$ _{-0.53 }^{+ 0.51}$	 \\\hline
$f_4^Z$                 &$ _{-0.96}^{+0.95}$&$ _{-0.67}^{+0.67}$ &$ _{-0.58 }^{+ 0.58}$&$ _{-0.44 }^{+ 0.45}$	 \\\hline
$f_5^Z$                 &$ _{-0.98}^{+0.95}$&$ _{-0.69}^{+0.68}$ &$ _{-0.57 }^{+ 0.57}$&$ _{-0.45 }^{+ 0.43}$	 \\\hline
\end{tabular*}
\end{table}
 A likelihood-based analysis using the total $\chi^2$  with the MCMC method is done by 
varying all the parameters simultaneously to extract 
simultaneous limits on all the anomalous couplings for the four benchmark luminosity chosen. 
 The two dimensional marginalised contours
 at $95~\%$ C.L.   in the  $f_4^\gamma$ -$f_4^Z$ and $f_5^\gamma$ -$f_5^Z$ planes
 are shown  in  Fig.~\ref{fig:2dmcmcplot-sigwithpol} 
  for the four benchmark luminosities chosen, using 
the  cross section together with the polarization asymmetries, i.e, using ($\sigma$ + pol.).
 The  outer most contours are for ${\cal L}=35.9$ fb$^{-1}$ and 
 the innermost contours are for ${\cal L}=1000$ fb$^{-1}$.
 The corresponding simultaneous limits on the aTGC couplings for four benchmark 
luminosities are presented in Table~\ref{tab:simul-limits}. The simultaneous limits are usually
less tight than the one-dimensional limits, but find the opposite in some case, which can
be seen comparing Table~\ref{tab:simul-limits}
with  Table~\ref{tab:single-limits}. 
The reason for this is the following. The cross section, the dominant observable, has a very little
linear dependence, while it has a large quadratic dependence on the couplings (see
 Eq.~(\ref{eq:sigma-1TeV})). As a result, when one obtains the limit on one parameter
 in the multi-parameter analysis, a slight deviation on any other parameter from zero 
 (SM point)  tightens the limit on the former  coupling.
 
\subsection{Role of polarization asymmetries in parameter extraction}
\begin{figure}
	\centering
	\includegraphics[width=0.24\textwidth]{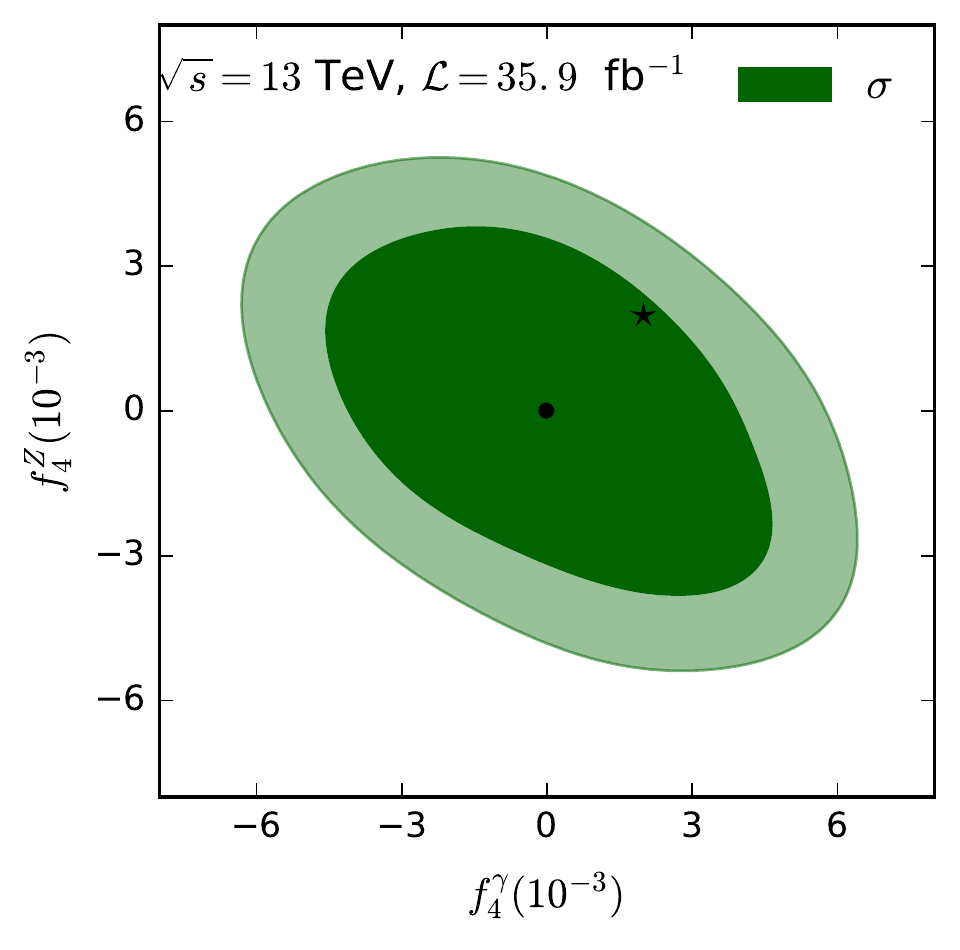}	
	\includegraphics[width=0.24\textwidth]{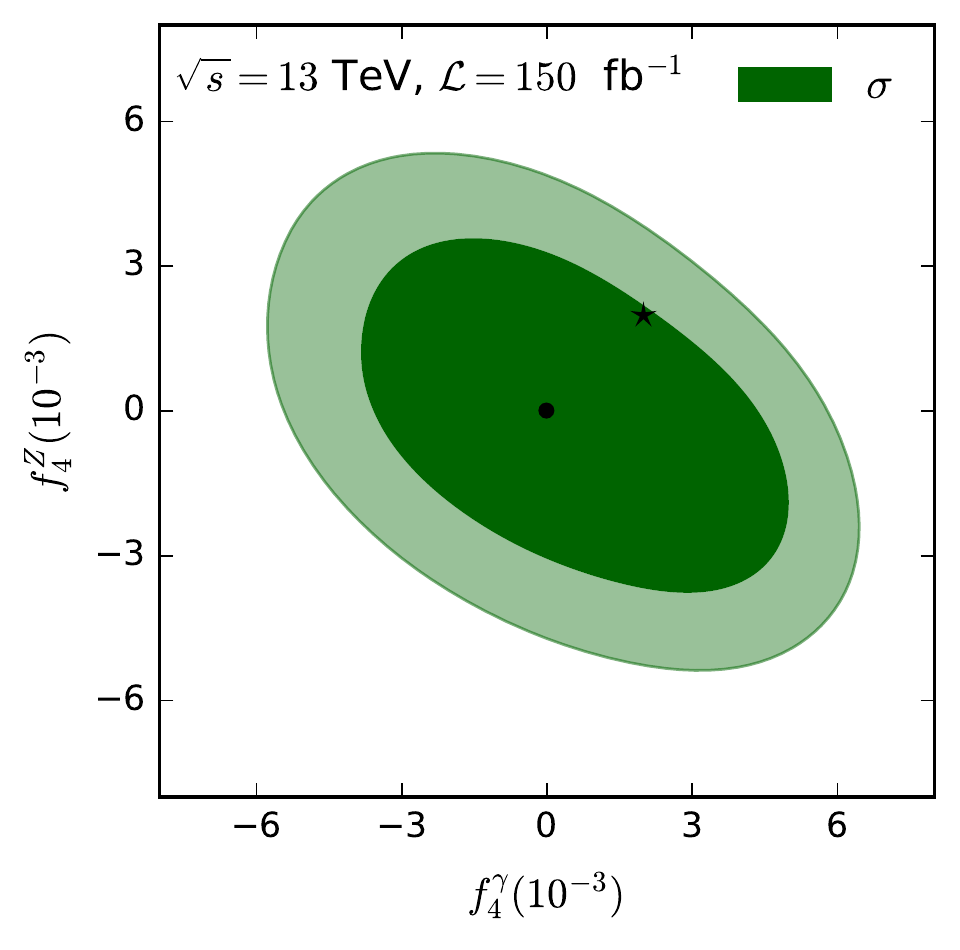}	
	\includegraphics[width=0.24\textwidth]{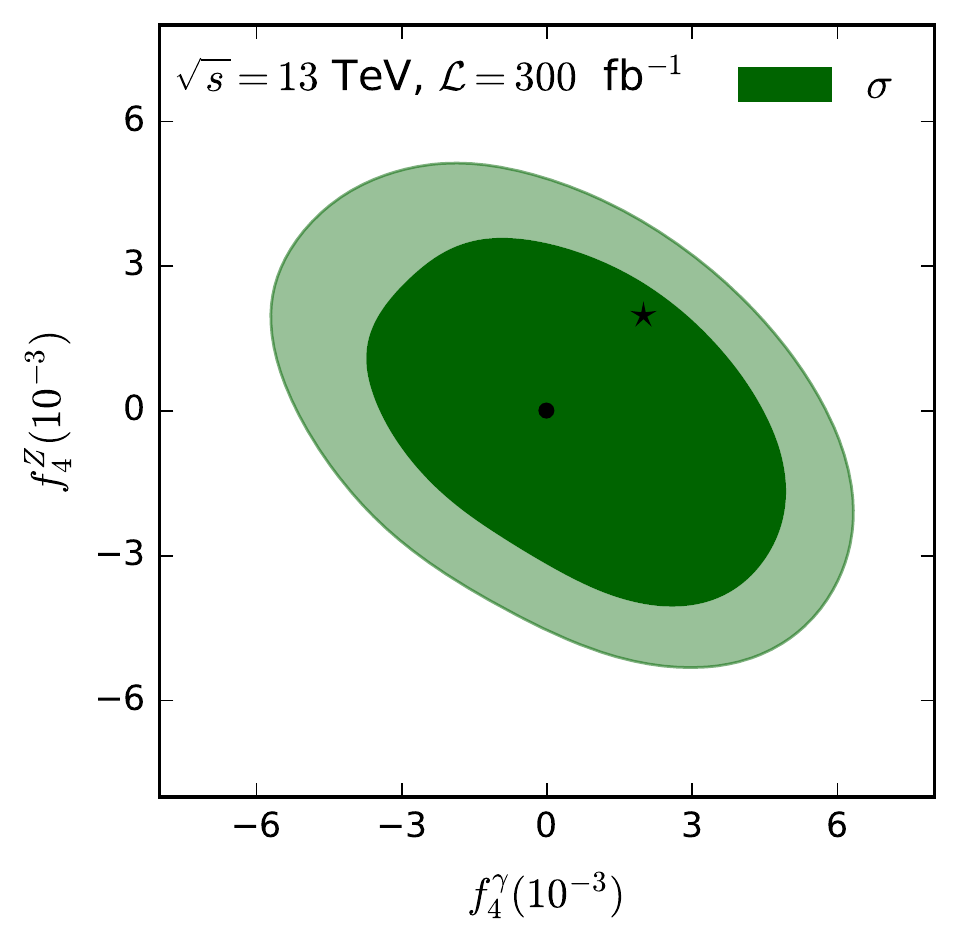}
	\includegraphics[width=0.24\textwidth]{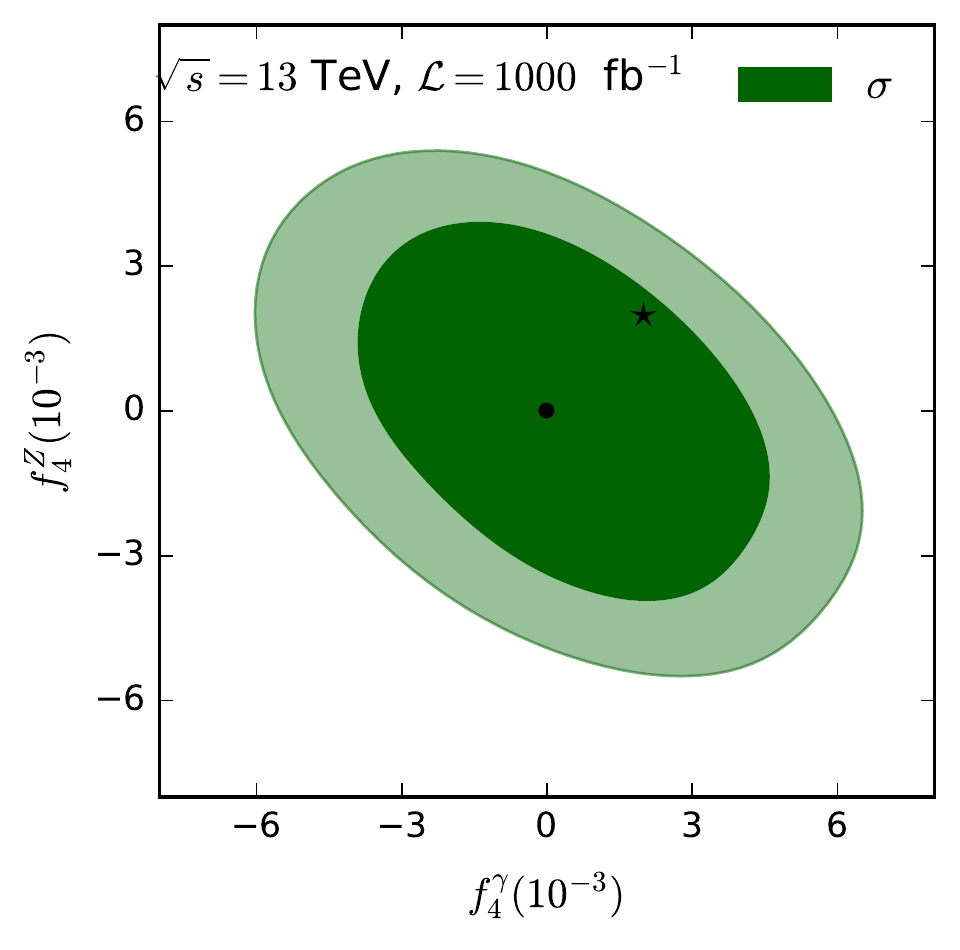}
	\includegraphics[width=0.24\textwidth]{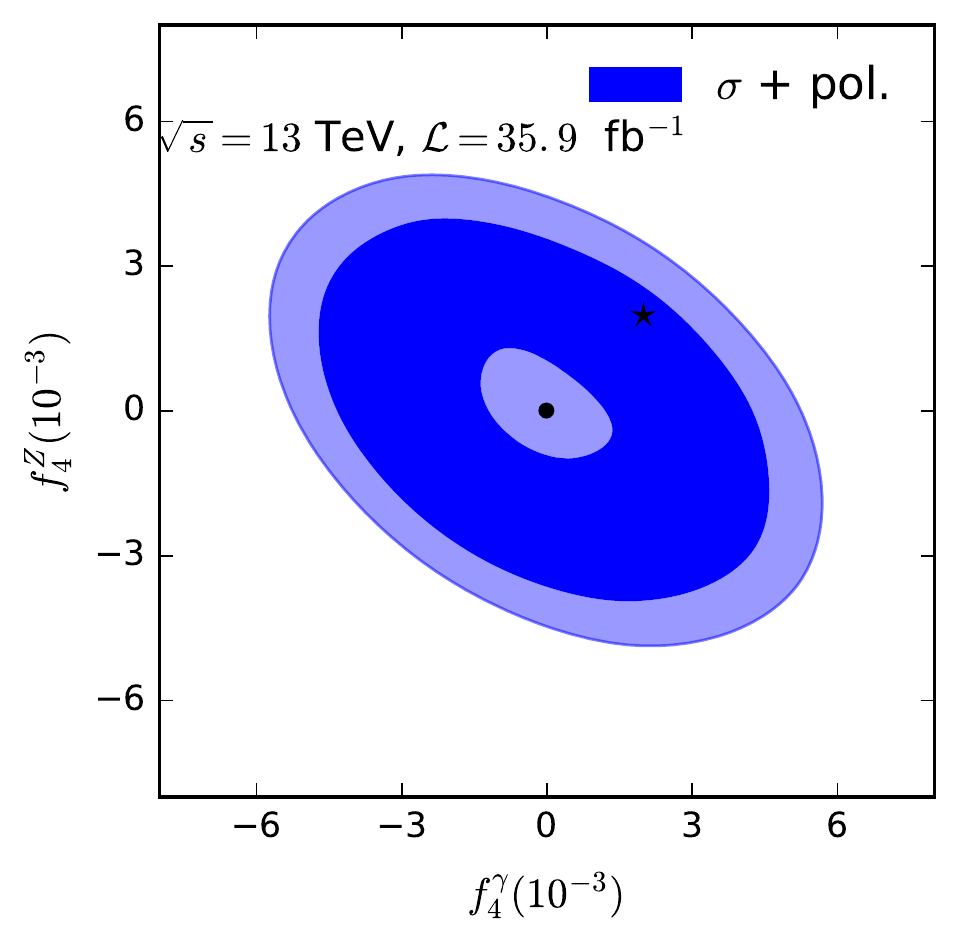}	
	\includegraphics[width=0.24\textwidth]{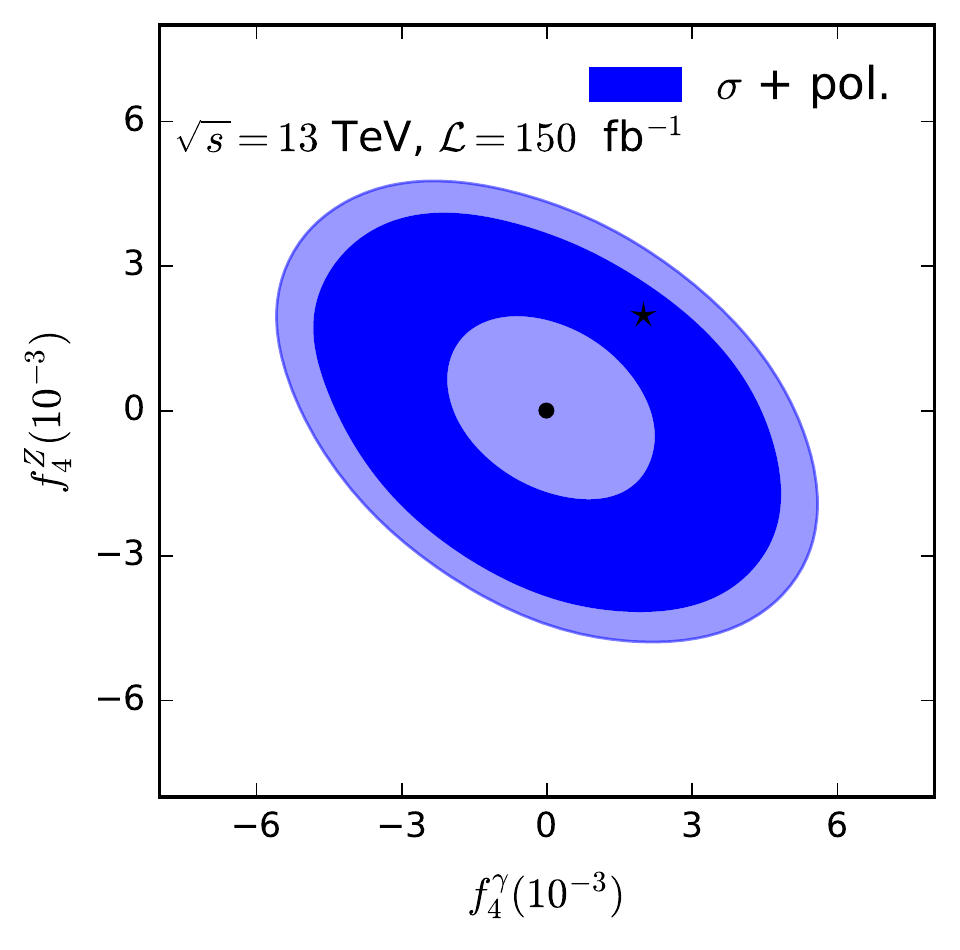}	
	\includegraphics[width=0.24\textwidth]{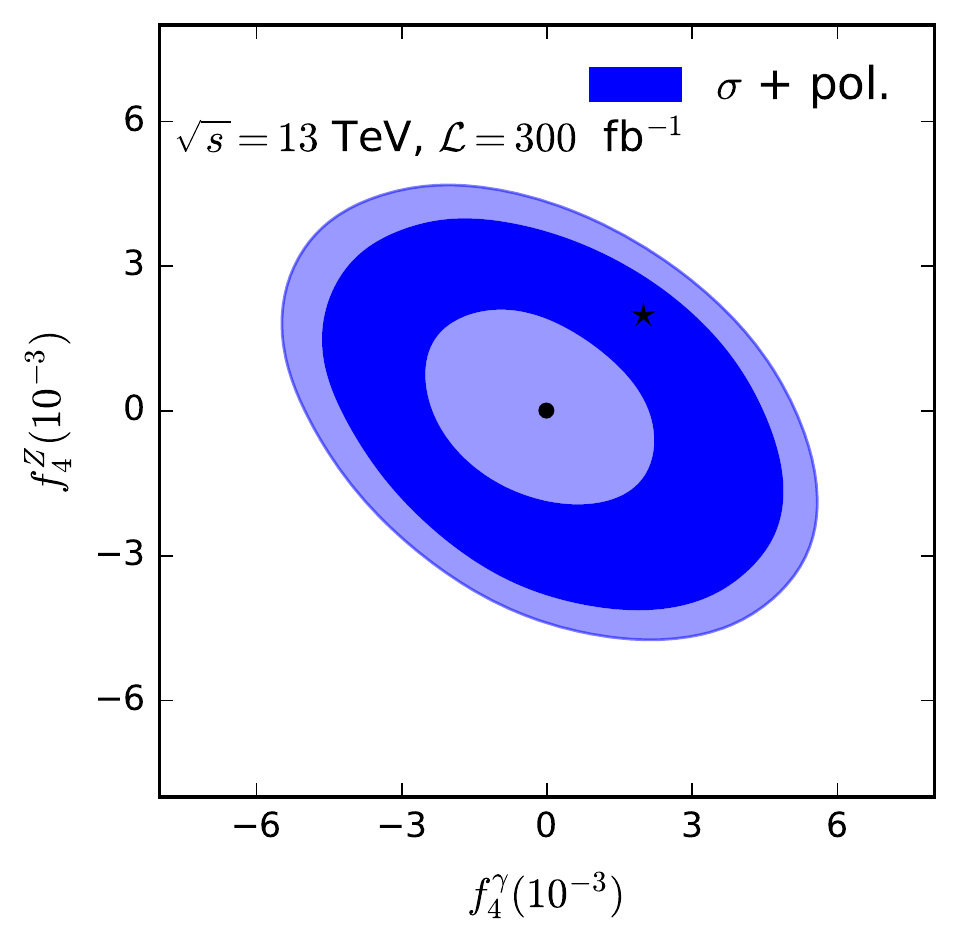}
	\includegraphics[width=0.24\textwidth]{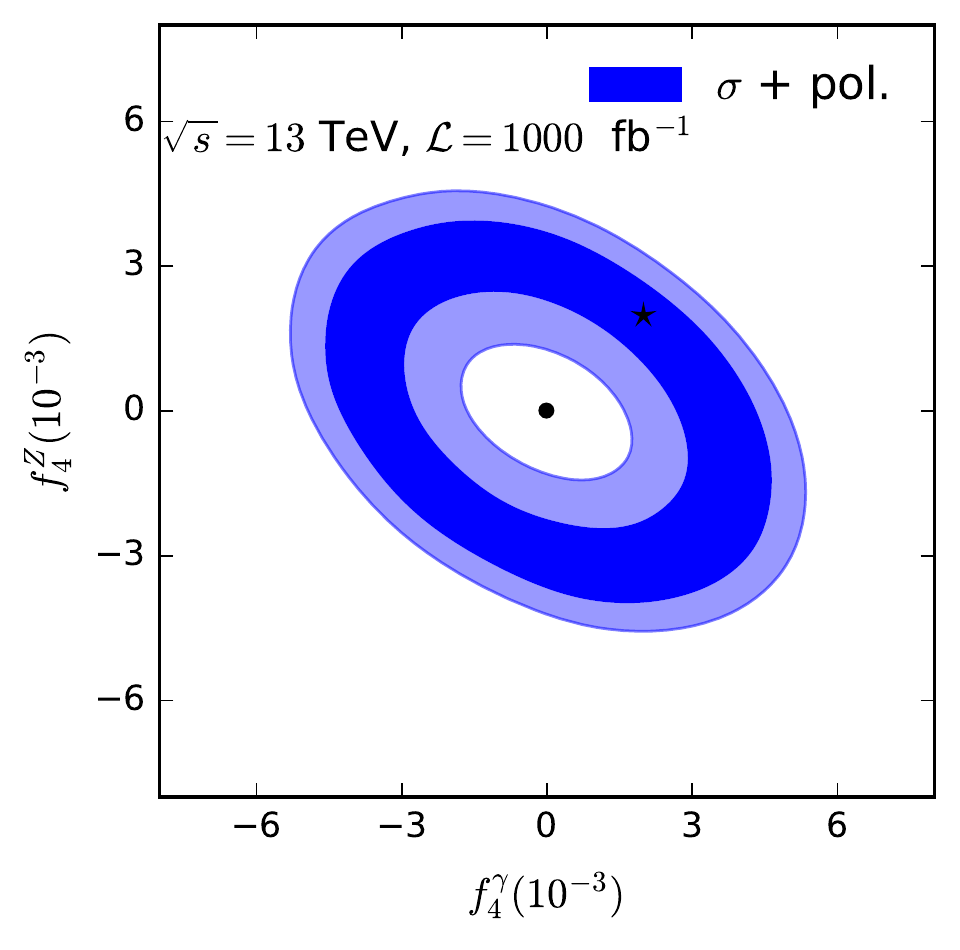}
	\includegraphics[width=0.24\textwidth]{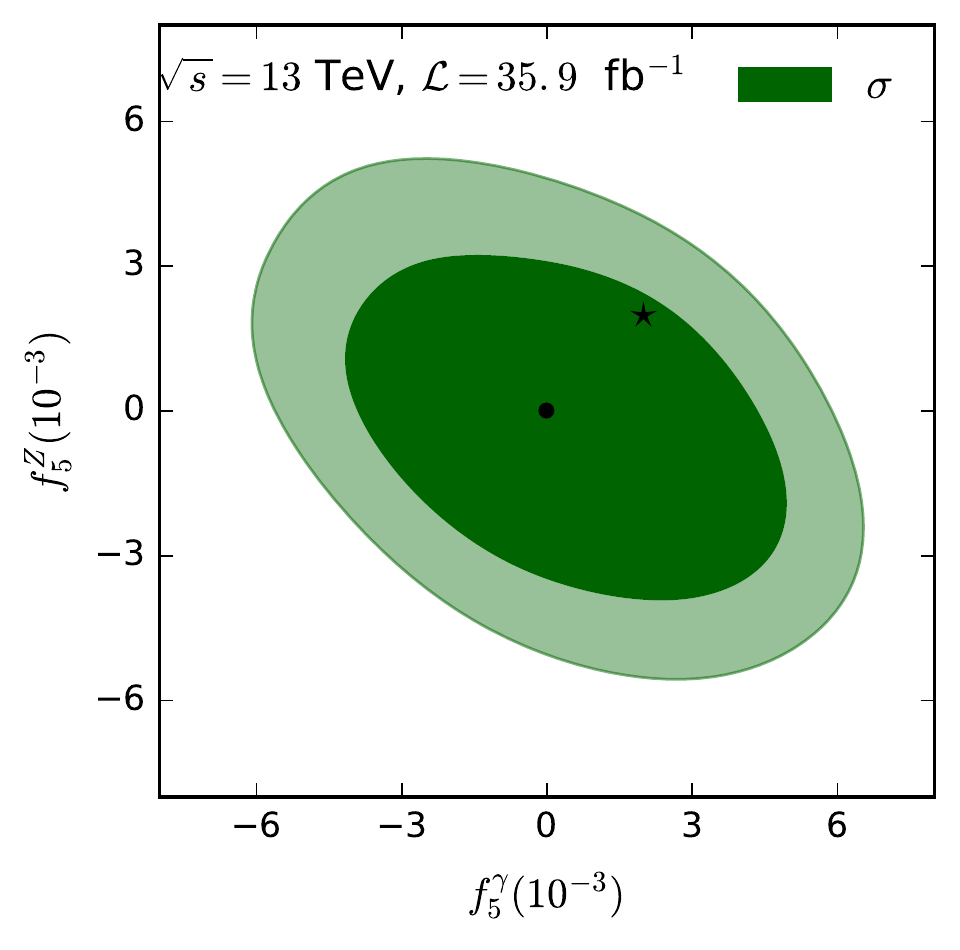}	
	\includegraphics[width=0.24\textwidth]{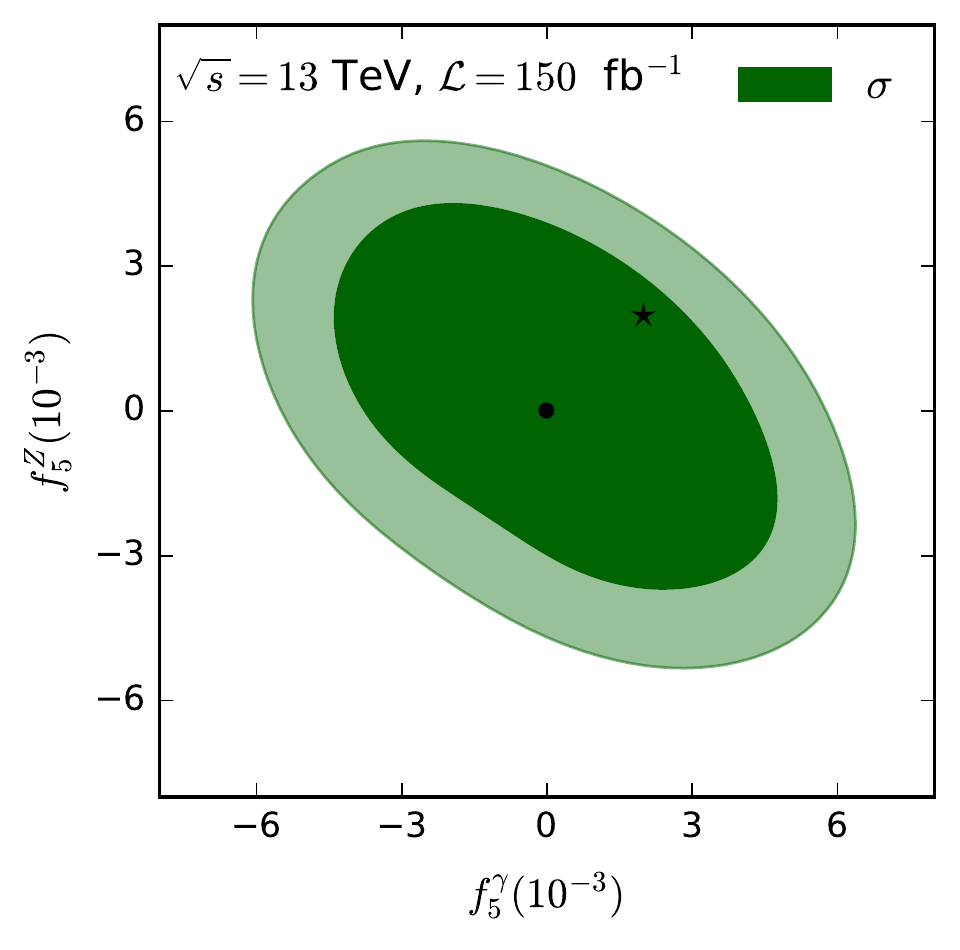}	
	\includegraphics[width=0.24\textwidth]{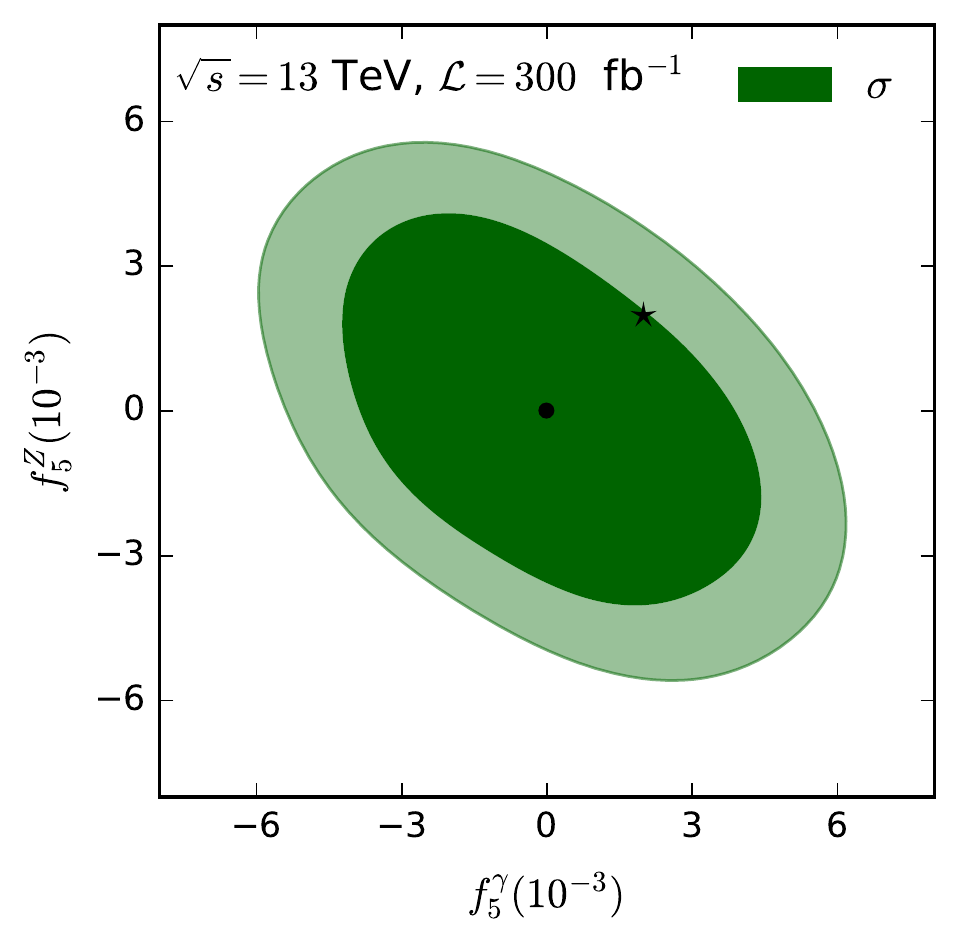}
	\includegraphics[width=0.24\textwidth]{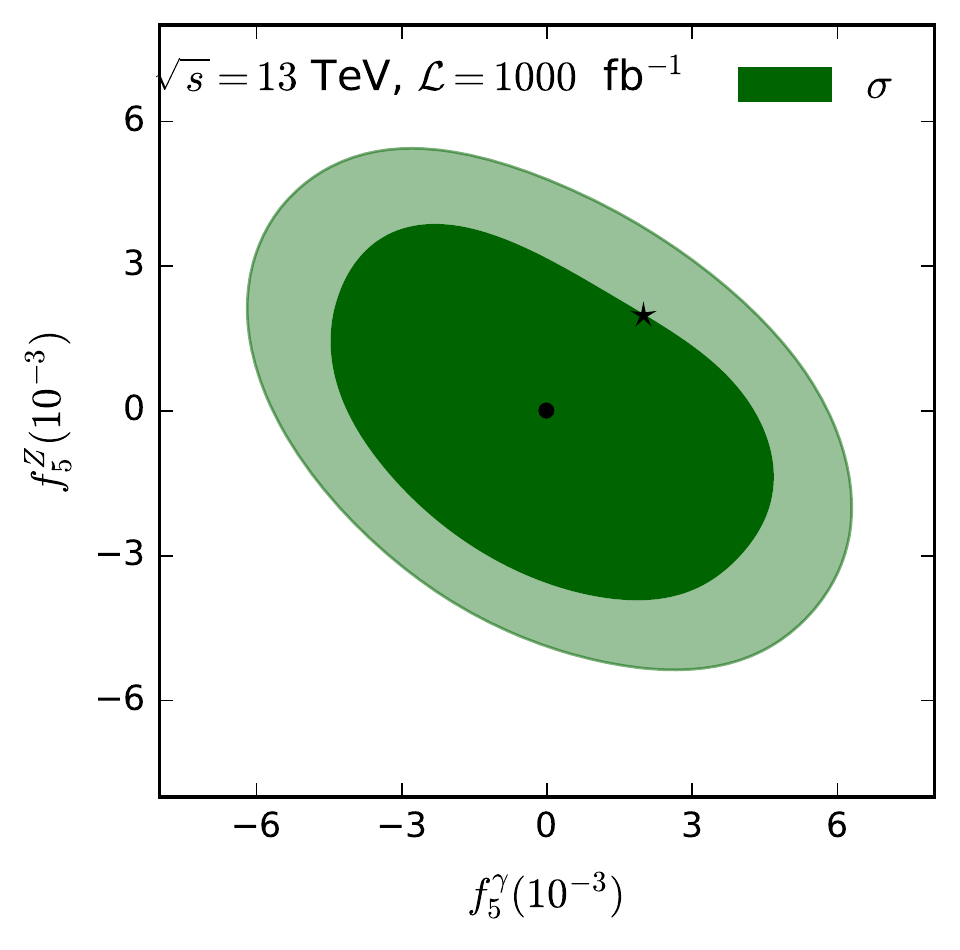}
	\includegraphics[width=0.24\textwidth]{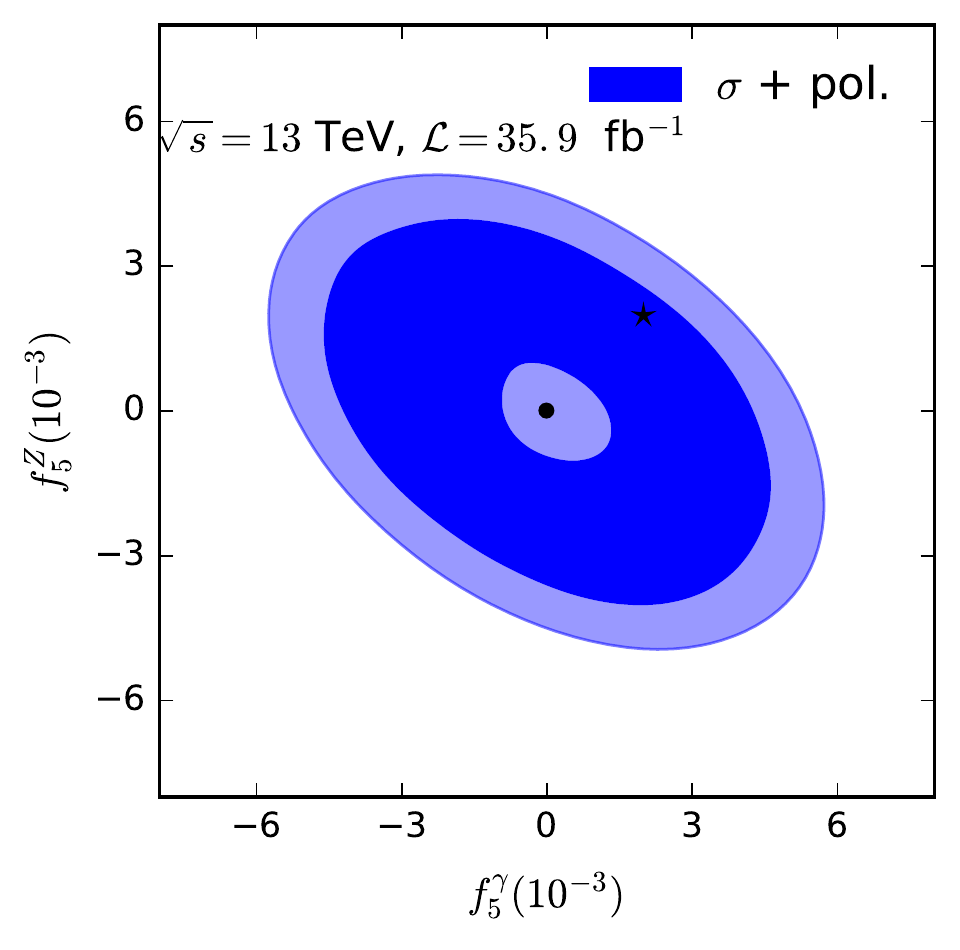}	
	\includegraphics[width=0.24\textwidth]{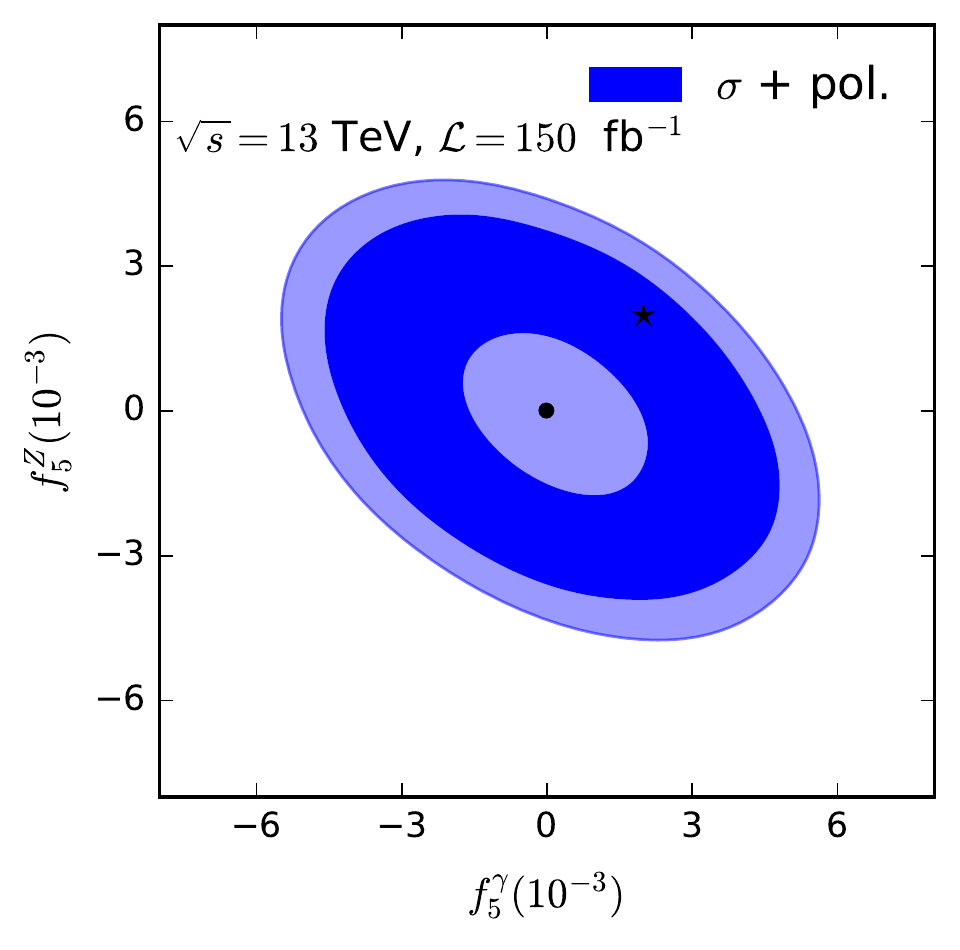}	
	\includegraphics[width=0.24\textwidth]{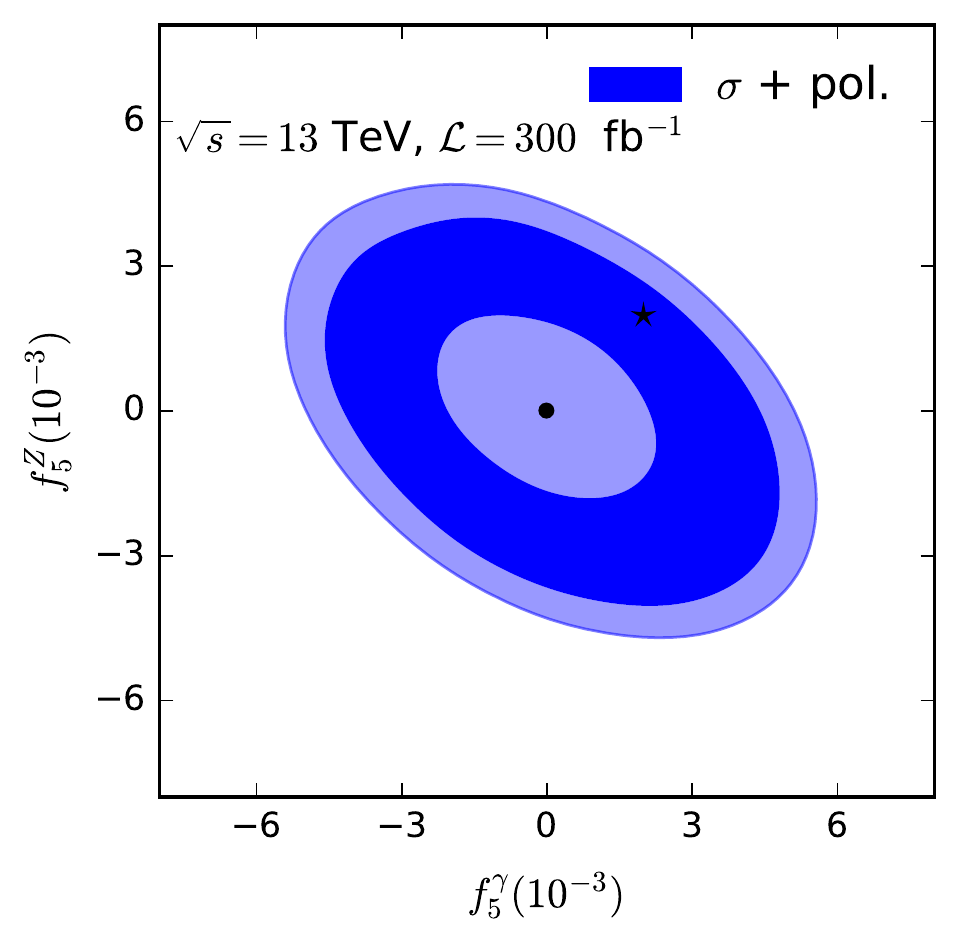}
	\includegraphics[width=0.24\textwidth]{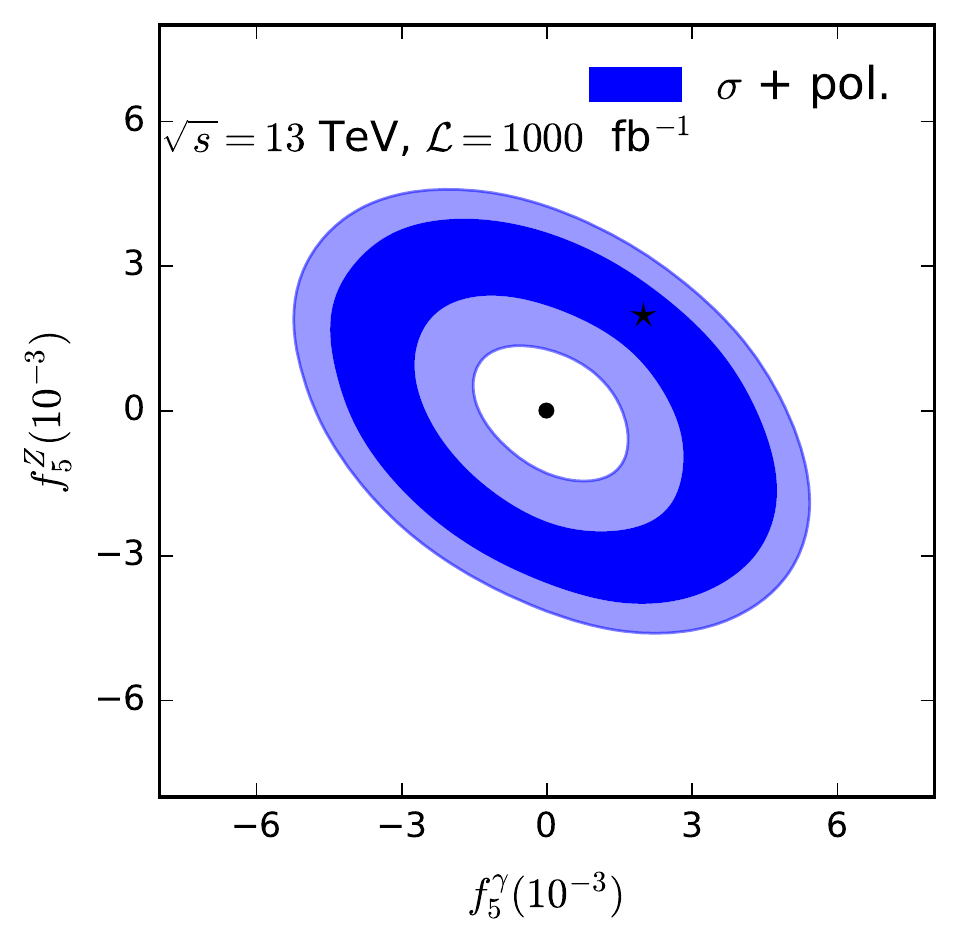}
	\caption{\label{fig:2dmcmcplot-bench2Lall} Comparison of $\sigma$ vs  
		($\sigma$ + pol.)  in two dimensional marginalised contours from MCMC for aTGC 
		benchmark $f_i^V=0.002$ in $f_4^\gamma$-$f_4^Z$ panel and $f_5^\gamma$-$f_5^Z$ panel   
		at $\sqrt{s}=13$ TeV for various luminosities in $ZZ$ production at the LHC. }
\end{figure}
The inclusion of polarization asymmetries with  the
cross section has no significant effect in constraining the anomalous couplings. 
The asymmetries  may still be useful  in extracting 
parameters if excess events were found at the LHC. 
To explore 
this, we do a toy analysis of parameter extraction using 
the data for all aTGC couplings $f_i^V=0.002$ (well above current limit)
and  use the MCMC method to extract back these parameters. 
In Fig.~\ref{fig:2dmcmcplot-bench2Lall}, we show two-dimensional marginalized contours
for the four benchmark luminosities for the benchmark aTGC couplings point $f_i^V=0.002$
in $f_4^\gamma$-$f_4^Z$ and $f_5^\gamma$-$f_Z^Z$ planes for the set of observables
$\sigma$ and ($\sigma$ + pol.) for comparison.
The {\em darker-shaded} regions are for $68~\%$ C.L., while {\em lighter-shaded} regions are for
$95~\%$ C.L. The dot ({\tiny $\bullet$}) and the star ($\star$) mark  in the plot are for the SM ($0,0$) and 
aTGC benchmark ($0.002,0.002$) points, respectively.
We note that the SM point is inside the $68~\%$ C.L. contours even at a high luminosity of 
${\cal L}=1000$ fb$^{-1}$  if we use only cross section as observable, see  row-$1$ and $3$
of Fig.~\ref{fig:2dmcmcplot-bench2Lall}. The distinction between the SM and the aTGC get improved when polarization asymmetries are included, 
i.e., the SM point is outside the $95~\%$ C.L. contour for luminosity of much less than  ${\cal L}=1000$ fb$^{-1}$, see row-$2$ and $4$
of the figure.
As the luminosity increases, from the left column to the right, the contours for ($\sigma$ + pol.) shrink around
the star ($\star$) mark maintaining the shape of a ring giving better exclusion of the SM  from aTGC benchmark. 
Polarization asymmetries are thus useful in the measurement of the anomalous couplings 
if excess events are found at the LHC.

\section{Conclusions}\label{sect:conclusion}

In conclusion, we studied anomalous triple gauge boson couplings
in the neutral sector in $ZZ$ pair production at the LHC and investigated
the role of $Z$ boson polarizations. The QCD correction in this process
is very high and can  not be ignored. We obtained the cross section and the asymmetries
at higher order in QCD. The aTGC contributes more in the higher $\sqrt{\hat{s}}$
region as they are momentum dependent. The major background $t\bar{t}Z+WWZ$  are
negligibly small and they vanish in the signal regions. Although the asymmetries
are not as sensitive as the cross section to the couplings, the asymmetry
$A_{x^2-y^2}$ is able to distinguish between $CP$-even and $CP$-odd couplings.
We estimate the one parameter as well as simultaneous limits on the couplings using all 
the observables   based on the total $\chi^2$ for luminosities $35.9$ fb$^{-1}$, $150$ 
fb$^{-1}$, $300$ fb$^{-1}$ and $1000$ fb$^{-1}$. Our one parameter limits are comparable to the best 
available limits obtained by the LHC~\cite{Sirunyan:2017zjc}.
The asymmetries are instrumental in extracting the parameters
should a deviation from the SM is observed at the LHC. We 
did a toy analysis of parameter extraction
with a benchmark aTGC coupling point with $f_i^V=0.002$ and found that polarization with the cross
section can exclude the SM from the aTGC point better than the cross section can do alone.
In this work, the observables for the aTGC are obtained at ${\cal O }(\alpha_s)$, 
while they are obtained in the next order in the SM. 
The NNLO result in aTGC, when available, is expected to improve the limits on 
the couplings.

\vspace{1cm}
\noindent \textbf{Acknowledgements:} We thank Dr. Debajyoti Choudhury, Dr. V. Ravindran, 
Dr. Ayres Freitas and Dr. Adam Falkowski for fruitful discussion. R.~R. thanks Department of Science 
and Technology, Government of India for support through DST-INSPIRE Fellowship 
for doctoral program, INSPIRE CODE IF140075, 2014.  

\appendixtitleon
\appendixtitletocon
\begin{appendices}
\section{Expressions of observables}\label{app:a}
\begin{eqnarray}\label{eq:sigma-1TeV}
\sigma (M_{4l} > 1~\text{TeV}) 
&=& 0.096685 
+ f_5^\gamma \times 1.9492 + f_5^Z \times 1.7106 \nonumber\\
&+& f_4^\gamma f_4^Z\times 51788 + f_5^\gamma f_5^Z\times 51204 
+ (f_4^\gamma)^2 \times 54933 \nonumber\\
&+& (f_4^Z)^2 \times 75432 
+ (f_5^\gamma)^2 \times 54507 + (f_5^Z)^2 \times 74466 \hspace{0.2cm}\text{fb}\nonumber\\
\end{eqnarray}
\begin{eqnarray}\label{eq:sigma-300GeV}
\sigma (M_{4l} > 0.3~\text{TeV}) 
&=& 7.9503 
+ f_5^\gamma \times 16.886 + f_5^Z \times 4.0609 \nonumber\\
&+& f_4^\gamma f_4^Z\times 58561 + f_5^\gamma f_5^Z\times 54131 
+ (f_4^\gamma)^2 \times 58771 \nonumber\\
&+& (f_4^Z)^2 \times 81647 
+ (f_5^\gamma)^2 \times 55210 + (f_5^Z)^2 \times 78325 
\hspace{0.5cm}\text{fb}\nonumber\\
\end{eqnarray}
\begin{eqnarray}\label{eq:sigma-700GeV}
\sigma (M_{4l} > 0.7~\text{TeV}) 
&=& 0.37616 
+ f_5^\gamma \times 3.8161 + f_5^Z \times 2.9704 \nonumber\\
&+& f_4^\gamma f_4^Z\times 55005 + f_5^\gamma f_5^Z\times 52706 
+ (f_4^\gamma)^2 \times 57982 \nonumber\\
&+& (f_4^Z)^2 \times 80035 
+ (f_5^\gamma)^2 \times 57131 + (f_5^Z)^2 \times 78515 \hspace{0.3cm}\text{fb}\nonumber\\
\end{eqnarray}
\begin{eqnarray}\label{eq:axz-300GeV}
\wtil{A_{xz}^{\text{num.}}} (M_{4l} > 0.3~\text{TeV}) 
&=& -0.77152 
+ f_5^\gamma \times 6.1912+ f_5^Z \times 7.8270 \nonumber\\
&+& f_4^\gamma f_4^Z\times 2869.5 + f_5^\gamma f_5^Z\times 396.94 
+ (f_4^\gamma)^2 \times 1029.7 \nonumber\\
&+& (f_4^Z)^2 \times 2298.7 
- (f_5^\gamma)^2 \times 274.02\nonumber\\
& -& (f_5^Z)^2 \times 1495.5\hspace{0.2cm}\text{fb}
\end{eqnarray}
\begin{eqnarray}\label{eq:axxyy-700GeV}
A_{x^2-y^2}^{\text{num.}} (M_{4l} > 0.7~\text{TeV}) 
&=& -0.04295 
+ f_5^\gamma \times 1.5563+ f_5^Z \times 0.37094 \nonumber\\
&+& f_4^\gamma f_4^Z\times 3299.8 - f_5^\gamma f_5^Z\times 5853.9 
+ (f_4^\gamma)^2 \times 4241.8 \nonumber\\
&+& (f_4^Z)^2 \times 5679.3 
- (f_5^\gamma)^2 \times 6520.1\nonumber\\
 &-& (f_5^Z)^2 \times 8559.3 \hspace{0.2cm}\text{fb}
\end{eqnarray}
\begin{eqnarray}\label{eq:azz-700GeV}
A_{zz}^{\text{num.}} (M_{4l} > 0.7~\text{TeV}) 
&=& 0.048175 
- f_5^\gamma \times 0.12125 - f_5^Z \times 1.5339 \nonumber\\
&-& f_4^\gamma f_4^Z\times 6449.2 - f_5^\gamma f_5^Z\times 5860.4 
- (f_4^\gamma)^2 \times 6344.7 \nonumber\\
&-& (f_4^Z)^2 \times 8907.4 
- (f_5^\gamma)^2 \times 6457.7\nonumber\\
& -& (f_5^Z)^2 \times 8346.8 
\hspace{0.2cm}\text{fb}
\end{eqnarray}

The asymmetries will be given as,
\begin{align}
\wtil{A_{xz}}=\dfrac{\wtil{A_{xz}^{\text{num.}}} (M_{4l} > 0.3~\text{TeV})}{\sigma (M_{4l} > 0.3~\text{TeV}) },\nonumber\\
A_{x^2-y^2}=\dfrac{A_{x^2-y^2}^{\text{num.}} (M_{4l} > 0.7~\text{TeV}) }{\sigma (M_{4l} > 0.7~\text{TeV}) },\nonumber\\
A_{zz}=\dfrac{A_{zz}^{\text{num.}} (M_{4l} > 0.7~\text{TeV})}{\sigma (M_{4l} > 0.7~\text{TeV}) }.
\end{align}

\section{Note on linear approximation}\label{app:b}
We note that, the linear approximation of considering anomalous couplings will be valid if the quadratic 
contribution on the cross section will be much smaller than the linear contribution, i.e.,
\begin{equation}
|f_i \times \sigma_i |\gg |f_{i}^2\times \sigma_{ii}|,~~~\text{or}~~ |f_i|\ll \frac{\sigma_i}{\sigma_{ii}},
\end{equation}
where $\sigma_{i}$ and $\sigma_{ii}$ are the linear and quadratic
coefficient of the coupling $f_i$ in the cross section.
Based on $\sigma (M_{4l} > 1~\text{TeV})$ in Eq.~(\ref{eq:sigma-1TeV}) the linear approximation constrain $f_5^V$ as
\begin{eqnarray}
|f_5^Z|\ll 2.2\times 10^{-5},~~ |f_5^\gamma|\ll 3.5\times 10^{-5},
\end{eqnarray}
which are much much smaller than the limit (see Eq.~(\ref{eq:CMS-limit})) 
observed at the LHC~\cite{Sirunyan:2017zjc}. To this end we
keep terms upto quadratic in couplings in our analysis. 
\end{appendices}

\bibliography{zpolzz_lhc}
\bibliographystyle{utphys}
\end{document}